\documentclass[reprint,
 amsmath,amssymb,
 aps,
pra,
]{revtex4-2}

\usepackage[english]{babel}
\usepackage{graphicx}
\usepackage{dcolumn}
\usepackage{bm}

\usepackage{amsmath, braket} 
\usepackage{amssymb}
\usepackage{xcolor}
\usepackage{bbold}
\usepackage{hyperref}
\usepackage{comment}
\usepackage[caption=false]{subfig}
\newcommand{\op}[1]{\hat{#1}}

\begin{document}
\preprint{APS/123-QED}

\title{Pairwise Measurement Induced Synthesis of Quantum Coherence}
\author{Mariia Gumberidze}
\email{gumberidze@optics.upol.cz}
 \author{Michal  Kol\'a\v r}
 \email{kolar@optics.upol.cz}
\author{Radim Filip}
\email{filip@optics.upol.cz}
\affiliation{
Department of Optics, Palack\'y University, 17. listopadu 12, 771 46 Olomouc, Czech Republic
}

\begin{abstract} 
Quantum coherent superpositions of states with different energies, i.e., {\it states with coherence} with respect to energy basis, are important resource for modern quantum technologies. States with {\it small} coherence can be obtained either autonomously, due to the effect of a weak coherent drive or, potentially, due to the presence of an environment. In this paper, we propose a measurement-based protocol for quantum coherence synthesis from individual systems (with  low initial coherence) into a global (and higher) coherence of the joint system. As an {\it input}, it uses $N$ non-interacting copies of two-level systems (TLS), with {\it low} initial energy and coherence. These can be supplied by, e.g., a weak external drive or can result from an interaction with a bath. This protocol {\it conditionally} synthesizes an output state with higher energy and coherence than the initial state had, representing an universal process whose rules have not been  well studied, yet. In addition to energy and coherence, we study the quantity called mutual coherence, showing increase after the protocol application, as well. This approach is based on application of sequential {\it pairwise} projective measurements on TLS pairs (conditionally removing their ground states), that are diagonal in the TLS energy basis. The functionality of the coherence synthesis is robust with respect to dephasing effects of the TLS environment on the system. Our approach may show its benefits in quantum sensing, quantum batteries charging, or other applications where synthesis of a larger coherent system from smaller (weaker) resources is useful. 
\end{abstract}

\maketitle
\section{\label{intro} Introduction}
The fields of quantum information processing (QIP) \cite{SlussarenkoAPR2019,BruzewiczAPR2019,Wendin_2017}, quantum metrology (QM) \cite{DegenRevModPhys2017}, and quantum thermodynamics (QT) \cite{Anders_2017,Goold_2016} are rapidly developing in recent years. Their common denominator is the direct use of quantum coherence, i.e., the presence of the off-diagonal terms in the state description with respect to the relevant basis. In the realm of QIP the coherence is present typically in all processing stages. Also the first quantum simulators and quantum computer units profit from diverse quantum coherence of large systems. For QT it is still an open question if at all and eventually how to fully exploit the presence of coherence in the corresponding tasks \cite{StreltsovRevModPhys2017}. Irrespective of the particular area of coherence utilization, its manipulation (namely distillation) has been studied in recent years in deterministic \cite{LiuPRL2019,WuNPJ2020} as well as probabilistic (and another party-assisted) settings \cite{PangQIP2020,WuNPJ2020,Starek2021}, in a coherence resource theoretic framework.

As classical thermodynamics is historically motivated by study of {\it energy} transformations and its concentration, QT, as its modern offspring, also focuses at, in a sense, complementary question. If, and eventually how, are we able to create \cite{giacomoPRL2018,romanancheyta2020enhanced,ArchakNPJ2020,GUARNIERIPLA2020}, transfer, concentrate (from smaller elementary systems into a more complex ones), and protect \cite{Campaioli2020} quantum {\it coherence} as a potential thermodynamic resource \cite{ancheytaPRE2019,KlatzowPRL2019,StreltsovRevModPhys2017,kolarPRA2017}. Quantum thermodynamics studies among other topics the use of energy and, possibly, coherence stored, e.g., in quantum batteries \cite{alicki2013,binderNJP2015}, originally introduced as a set of $N$ two level systems. The research still continues with further investigation on their energy aging \cite{Pirmoradian2019}, stability \cite{santos2019}, and power enhancement \cite{campaioliPRL2017}.

We tackle the above mentioned question on transfer of coherence (always with respect to energy basis from now on) and its concentration in a bottom-top approach, building on our previous work \cite{Gumberidze2019}. It has introduced coherent quantum battery, conditionally synthesized from a pair of TLS copies (cells). This has been accomplished by means of projection-based, conditional charging protocol synthesizing the factorized coherent cells into a higher-dimensional, non-factorizable coherent battery and charging it in the sense of increasing its energy and coherence jointly. The coherence transformation has been gained by using {\it incoherent} global projector. The projector, removing the ground states of TLS, rendered itself to be {\it universal} in the sense of increasing energy, coherence, and mutual (correlated) coherence \cite{XiSciRep2015,GuoPhysRevA2017,WangSciRep2017,TanPhysRevLett2018,Kraft_2018} simultaneously, while its structure is independent on the input states \cite{Gumberidze2019}. Such projection was efficient for weakly (coherently) excited TLS.

In this paper we substantially increase the protocol feasibility by changing the universal projection-based protocol globally applied to the system of $N$ weakly coherent, non interacting TLS (inspired by \cite{binderNJP2015,alicki2013}, where all TLS are considered initially almost in their ground states \cite{Gumberidze2019}), to many pairwise projections on the same TLSs. Moreover, we focus on keeping the universal protocol character mentioned in the previous paragraph. These are applied in a sequential manner, synthesizing factorized TLS into a higher-dimensional coherent system, while increasing its energy and coherence jointly. The pairwise approach is experimentally accessible, as no complex, multi TLS projection is needed. Notably, the simultaneous energy and coherence (with respect to energy eigenbasis) increase, is universally gained by using diagonal pairwise projectors independently of the input weakly excited states. We consider the choice of energy eigenbasis as a natural reference for coherence evaluation, being inspired, e.g., by works on quantum batteries \cite{binderNJP2015,alicki2013}, which are originally the energy storage devices. As such, their quantum mechanical properties should be studied with respect to this basis as well. Other methods employ the projection-based manipulations as well, but utilize them for different purposes, such as, e.g., the stabilization process of an open quantum battery \cite{Campaioli2020}.

Our simultaneous study of energy and coherence gain is complemented by the interest in the behavior of {\it mutual (correlated) coherence} \cite{XiSciRep2015,GuoPhysRevA2017,WangSciRep2017,TanPhysRevLett2018,Kraft_2018}. 
The mutual coherence allows us to distinguish the contributions to the coherence from the global ($N$-partite) state of our system and the individual contributions originating from all local states, obtained by partial tracing-out of the remaining TLS. {\color{black}From a thermodynamic perspective, mutual coherence also quantifies the difference in work extracted from a given state globally vs. locally if certain thermodynamic process is performed \cite{kammerlander}.}
By virtue of using such quantity, we show that our conditional protocol is (in the case of successful outcome) capable of {\it simultaneous} increase of the system energy, coherence, and mutual coherence. Thus, it transforms the factorized initial state of the system into a more coherent, non-factorizable final state. The coherence gain scales (for low excitation) as an increasing logarithmic function of $N$, while energy increases with $N$ linearly for the pairwise protocol studied here.

We test as well, if the protocol retains its function while the effect of local dephasing reduces the coherence of the initial system state progressively. The same test is performed regarding the effect of dephasing on the final (resulting) state of the protocol. In both cases the results suggest that the protocol function sustains such effect of environment.

The paper is organized as follows. Sec.~\ref{pairwise} describes the pairwise protocol applied on the pure initial states of the TLS and the quantities of interest. In Sec.~\ref{dephasing} the consequences of dephasing effect on TLS before and after the protocol are described.
Conclusions and outlook are given in the last Sec.~\ref{conclusions}. 

\begin{figure}
\includegraphics[width=0.95\columnwidth]{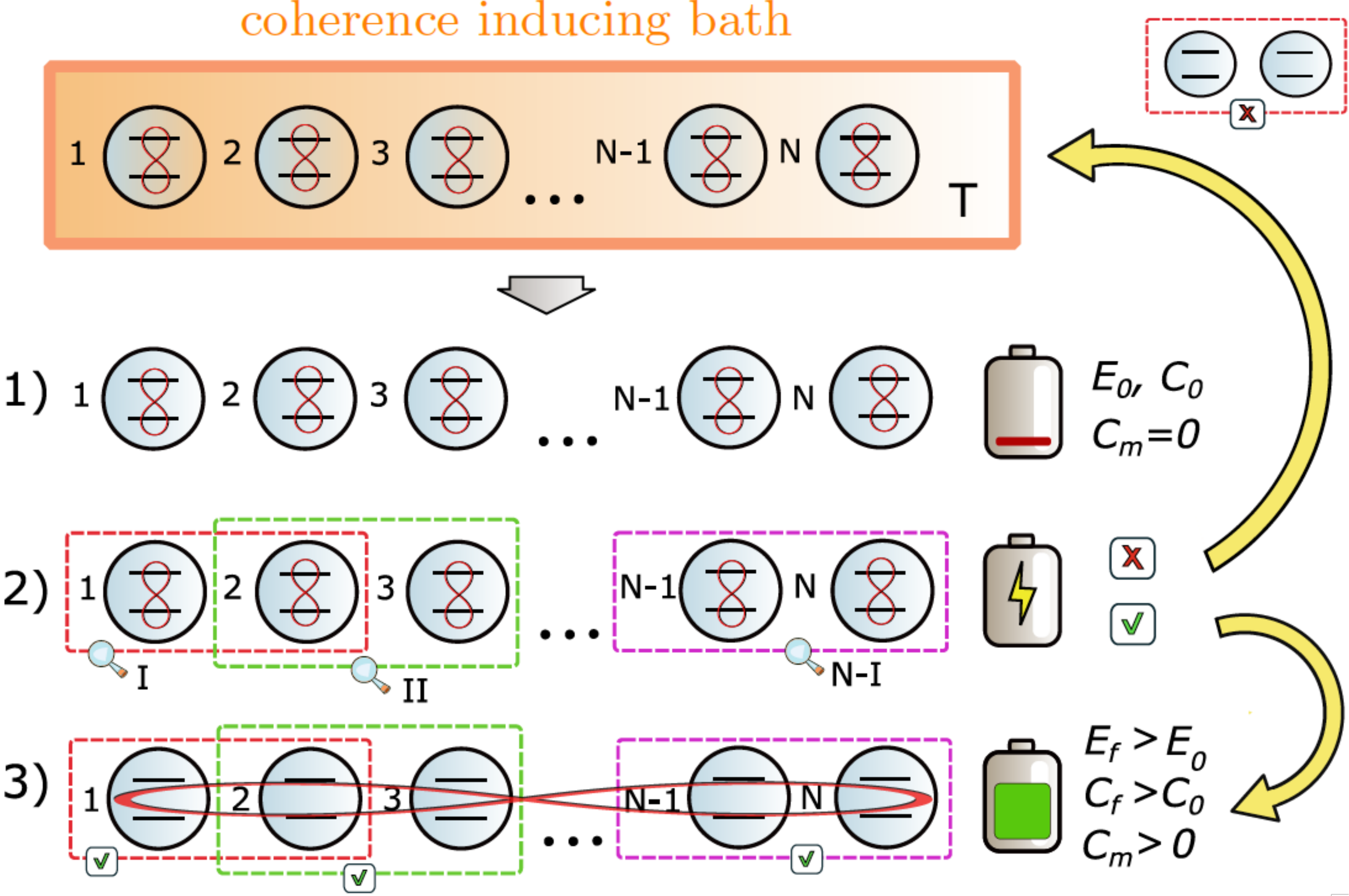}
\caption{Schematics showing the idea of $N$-TLS (two-level systems) coherence synthesis in three steps. In the initial {\bf step~1)} the TLS leave a low-coherence inducing bath \cite{giacomoPRL2018} and enter the process, assumed to be almost discharged with low initial energy $E_0^{(N)}$ and coherence $C_0^{(N)}$ (marked by red loops connecting the levels), superscript $N$ refers to the number of TLS used. {\bf Step~2)} represents the {\it conditional} stage of the synthesis operation. In this step the measurements on pairs of TLS in a sequence marked by $\{{\rm I,II},... \}$ are performed. If all measurements on pairs in this step are successful with the single run probability of success $p_s^{(N)}$, the coherence is synthesized and the system has higher energy $E_f^{(N)}>E_0^{(N)}$ and increased coherence $C_f^{(N)}>C_0^{(N)}$ as well, both with respect to the eigenbasis of Hamiltonian \eqref{eq-Ham-N}.  {\bf Step 3)} TLS constituting the system become correlated after the successful outcome of step 2 (represented by a large red loop in step 3). Moreover, the mutual coherence, Eq.~\eqref{eq-C-mut}, increases, $\Delta C_{m}^{(N)}>0$, during the protocol. If the pairwise measurements turned out to be unsuccessful in step 2, with the single run probability $p_f^{(N)}=1-p_s^{(N)}$, the system is completely discharged and incoherent. To achieve a successful synthesis, TLS can be recycled by bringing them in contact with the coherence-inducing bath again and the protocol is {\it repeated until success} (RUS).}
\label{scheme-2}
\end{figure}

\begin{figure}
\centering
\subfloat[\label{fig-delta-E}]{
\includegraphics[width=0.95\columnwidth]{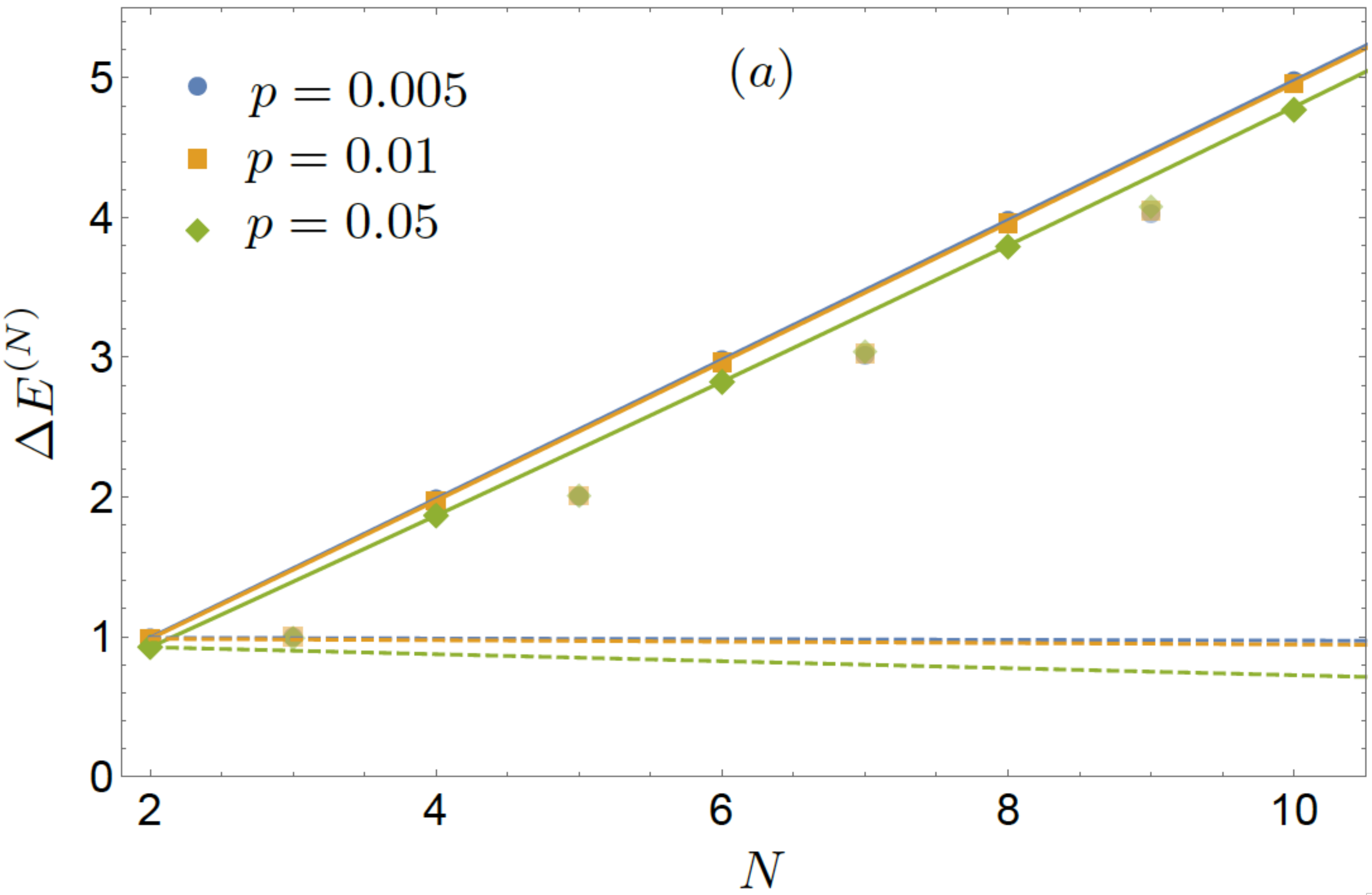}}
\hfill \vspace{-0.5cm}
\subfloat[\label{fig-delta-C}]{
\includegraphics[width=0.95\columnwidth]{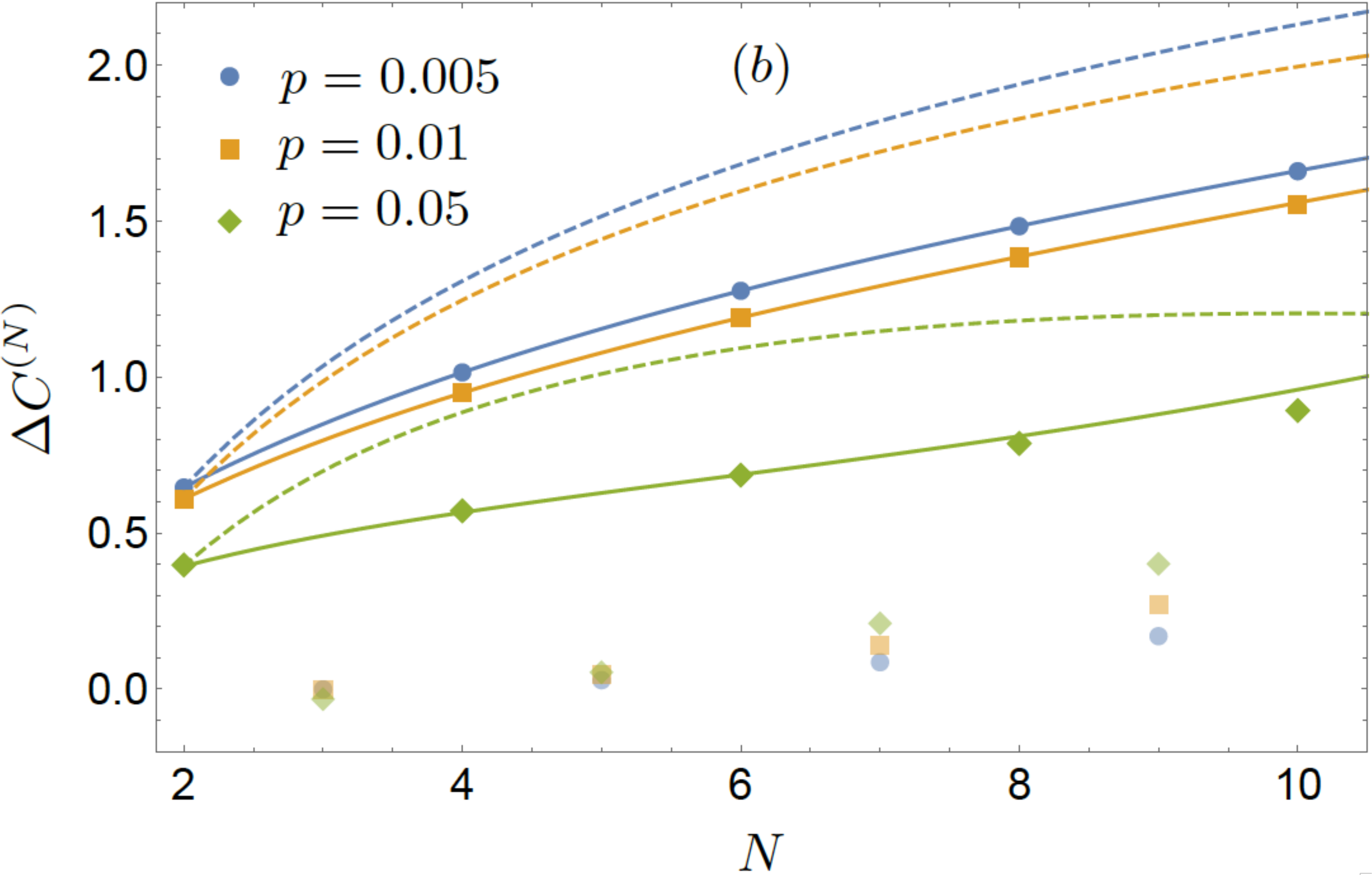}}
\hfill \vspace{-0.5cm}
\subfloat[\label{fig-delta-Cm}]{
\includegraphics[width=0.95\columnwidth]{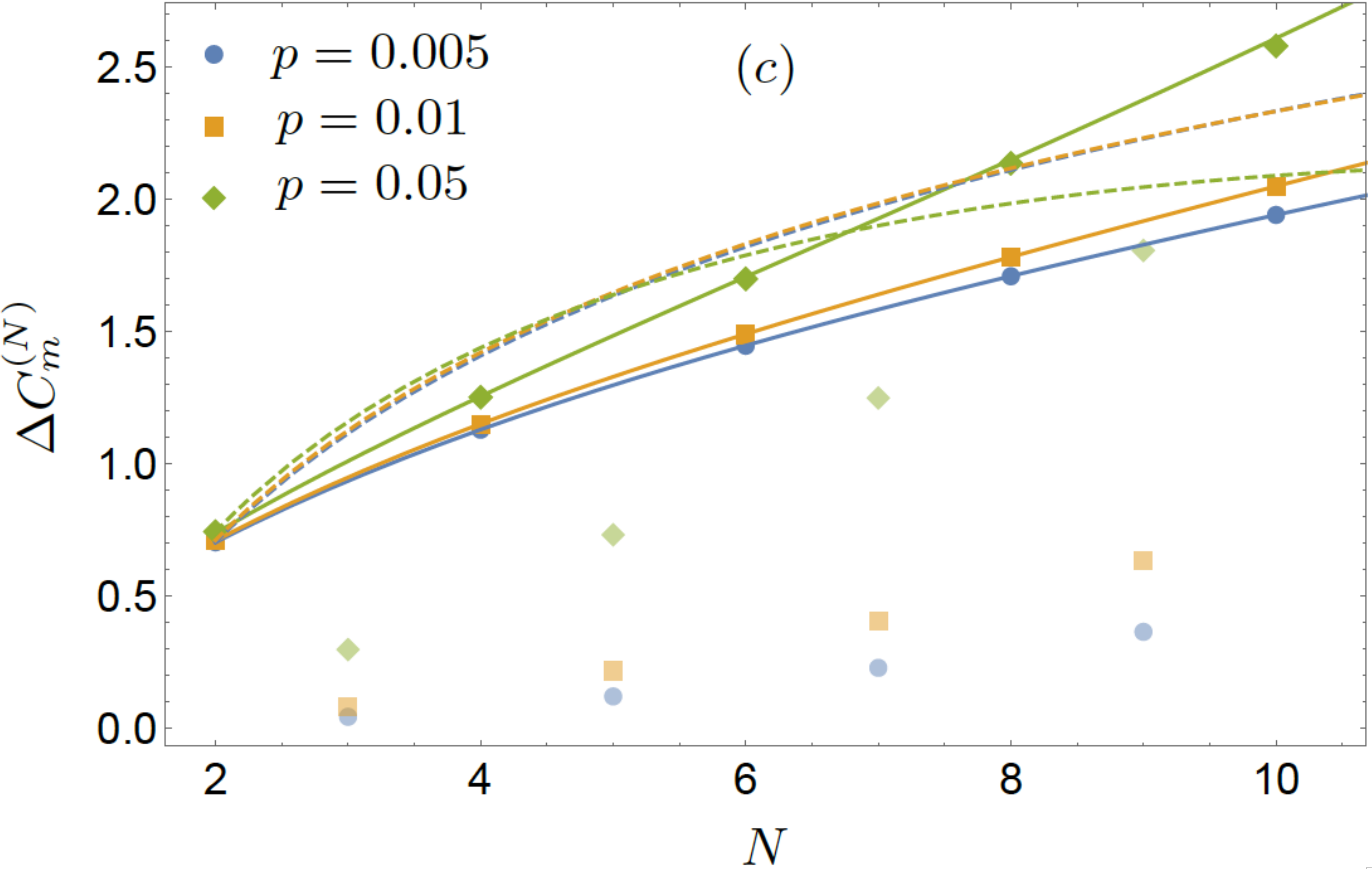}}
\vspace{-0.3cm}
\caption{Conditional synthesis of energy, coherence and mutual coherence of $N$ TLS. \protect\subref{fig-delta-E} The plot of the normalized average energy gain $\Delta E^{(N)}$, Eq.~\eqref{eq-pure-DE-N}, \protect\subref{fig-delta-C} the relative entropy of coherence gain $\Delta C^{(N)}$, Eq.~\eqref{eq-pure-DC-N}, and \protect\subref{fig-delta-Cm} the mutual coherence gain $\Delta C_{m}^{(N)}$, Eq.~\eqref{eq-C-mut},  of $N$ TLS in pure initial state, Eq.~\eqref{eq-pure-Psi-ini-N}. The values are plotted versus the number of TLS $N$ for different excitation probabilities $p$, Eq.~\eqref{eq-psi-pure}. The differently marked discrete points are exact numerical results, full lines are approximate results from Sec.~\ref{approximations}, and dashed lines remind of the results of global protocol \cite{Gumberidze2019}.
Note the reversed order of colors of full curves for $\Delta C_{m}^{(N)}$ and others, reflecting that $\Delta C_{m}^{(N)}$ is a non-decreasing function of $p\ll 1$, on contrary to $\Delta C^{(N)}$ and $\Delta E^{(N)}$, see app.~\ref{pure_states_append}. 
The coherence gain  $\Delta C^{(N)}$ differs for odd and even $N$ with a significant advantage for even $N$. Note, that in panel (a) the results for $p=0.005$ and $p=0.01$ almost coincide. In panel (c), the pairwise protocol overcomes the global one, for $N\gtrsim 6$ and $p\gtrsim 0.01$, corresponding to an untypical situation, cf. other panels.}
\end{figure}

\begin{figure}
\centering
\subfloat[\label{fig-prob-of-succ}]{
\includegraphics[width=0.95\columnwidth]{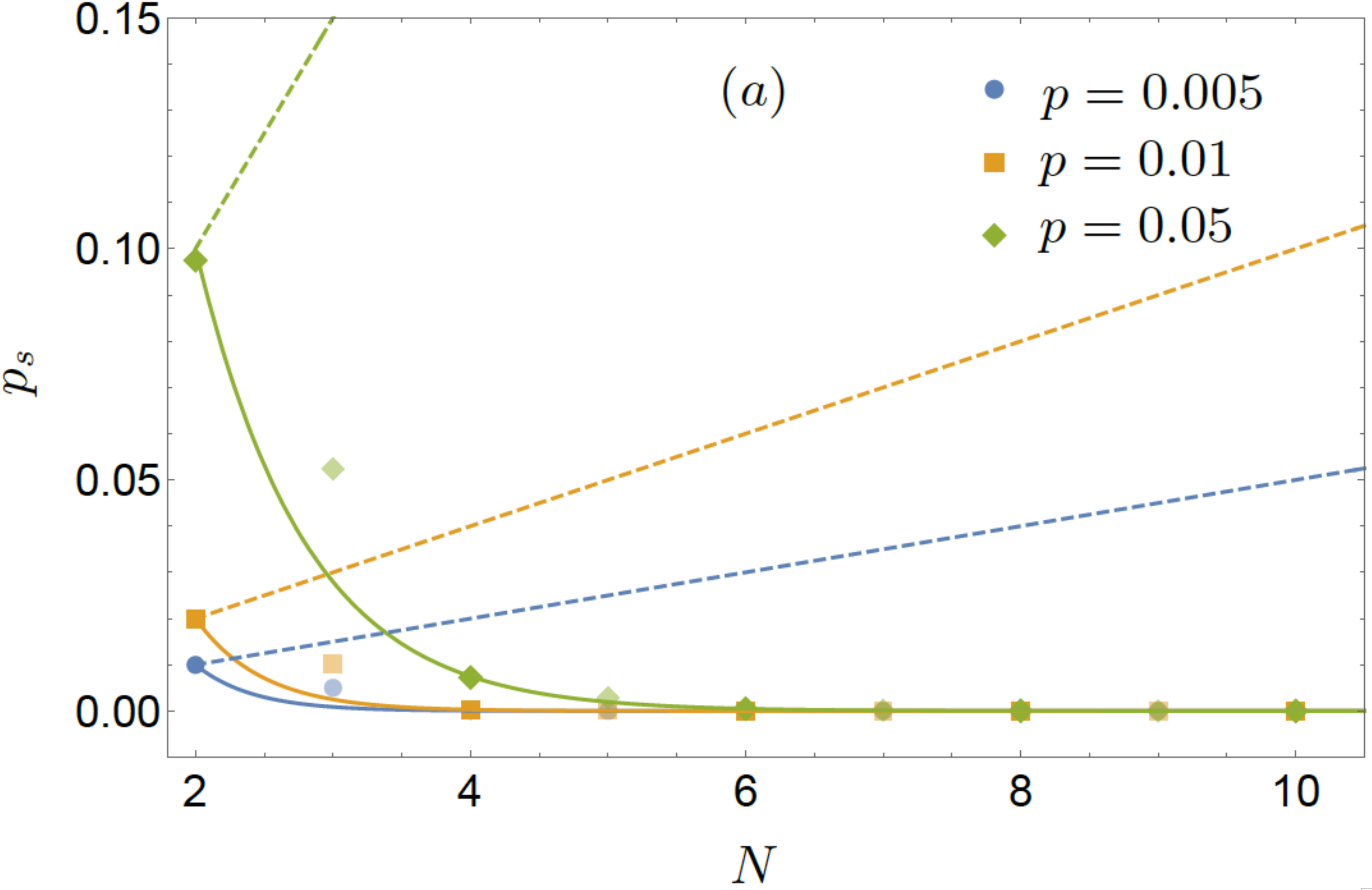}}
\hfill \vspace{-0.5cm}
\subfloat[\label{fig-prob-of-fail}]{
\includegraphics[width=0.95\columnwidth]{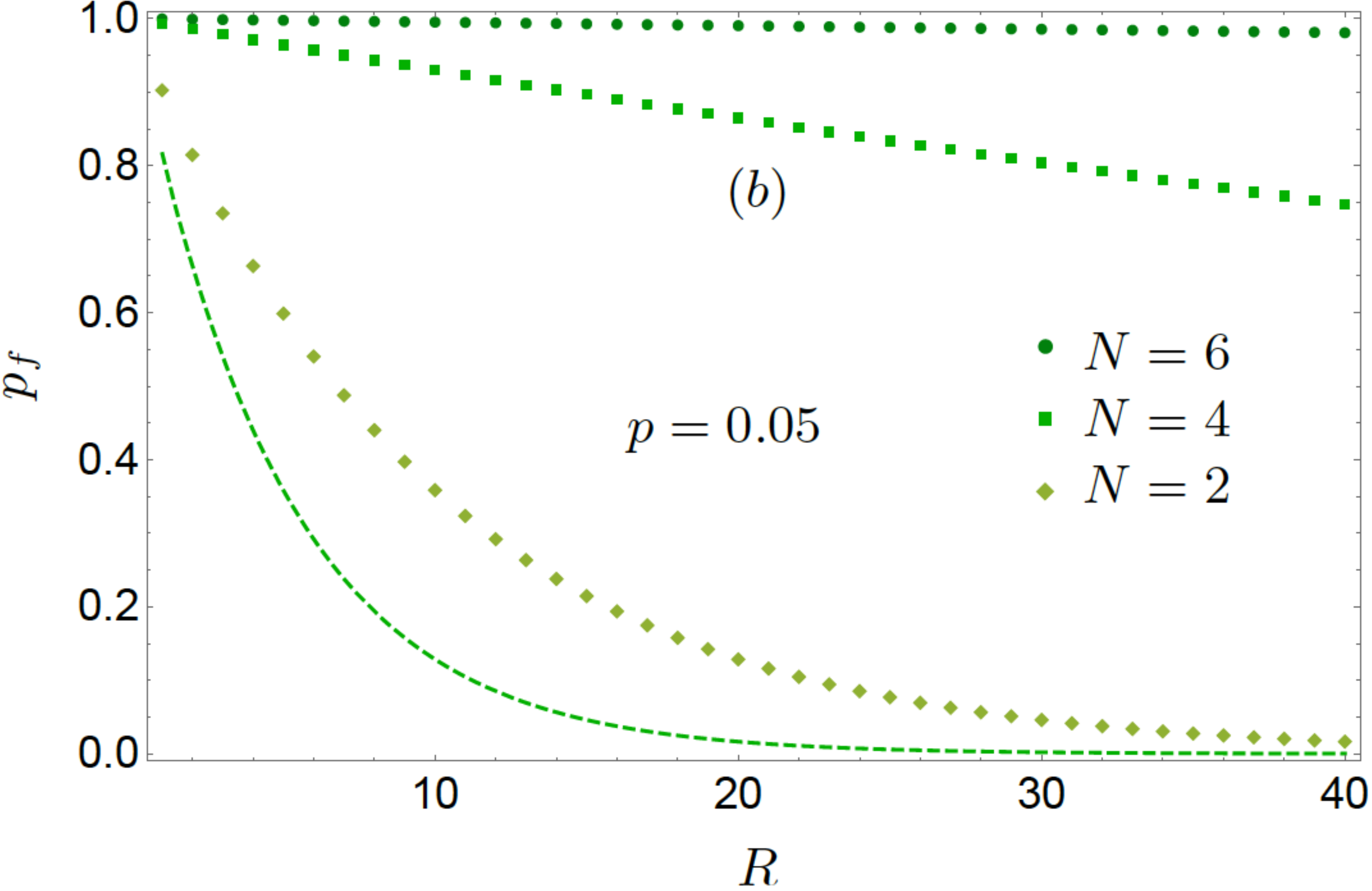}}
\vspace{-0.3cm}
\caption{Repeat-until-success strategy for coherence synthesis. \protect\subref{fig-prob-of-succ} The plot of probability of success $p_{s}$, Eq.~\eqref{eq-pure-succ-N}, of the projective measurement $\op{P}_1^{(N)}$, Eq.~\eqref{eq-proj-pair-total}, defined by the successful outcome with the final state $\ket{\Psi_f^{(N)}}$, Eq.~\eqref{eq-pure-Psi-fin-N}. The values are plotted versus the number of TLS $N$ for different excitation probabilities $p$, Eq.~\eqref{eq-psi-pure}. The probability of success $p_{s}\rightarrow 0$ with increasing $N$. Solid and dashed lines represent approximate expressions, see Sec.~\ref{approximations}, for even numbers $N$ of TLS for the pairwise and global \cite{Gumberidze2019} protocol, respectively. The discrete marked points represent exact numerical results for the pairwise protocol. \protect\subref{fig-prob-of-fail} The dependence of probability of failure $p_{f}$ in the repeat until success (RUS) strategy, see end of sec.~\ref{pairwise_protocol}, on the number of repetitions $R$ for $N=\{2,4,6\}$ TLS. The values are plotted for an example of the fixed probability of excited state $p=0.05$. On contrary to the probability of success $p_{s}$, probability of failure $p_{f}\rightarrow 1$ with larger number $N$ and smaller excitation probability $p$. The dashed line represents respective probability of failure of global protocol \cite{Gumberidze2019} for $N=4$ TLS, described in Sec.~\ref{approximations}.}
\end{figure}


\section{\label{pairwise} Pairwise protocol: measuring TLS in pairs}

\subsection{\label{setup} System and description} 

This section introduces the system of interest. It consists of $N$ independent (and noninteracting) copies of two-level systems (TLS), labeled $1,\dots ,N$ with a non-zero and constant energy gap $E$ between the ground and excited states. The preferred basis is given by the TLS's enegy eigenstates \cite{Gumberidze2019} labeling the ground and excited states, respectively, ${\ket{g_j},\ket{e_j}}$ for each TLS $j=1,\dots,N$.

The Hamiltonian of a single TLS
\begin{equation}
\op{H}_j=\frac{E}{2}\left(\ket{e_j}\bra{e_j}-\ket{g_j}\bra{g_j}\right),\quad j=1,\dots,N,
\label{eq-Hams}
\end{equation}
yields the Hamiltonian of the total system
\begin{equation}
\label{eq-Ham-N}
\hat{H}^{(N)}= \sum_{j=1}^{N} \op{H}_{j} \, \displaystyle{\bigotimes_{\substack{k=1\\k\neq j}}^{N} \op{\mathbb{1}}^{(k)}}.
\end{equation}

The average energy and coherence of the system are defined as
\begin{equation}
\langle E\rangle={\rm Tr}(\op{\rho}\op{H}), \quad C(\hat{\rho})\equiv S(\hat{\rho}||\hat{\rho}^{diag})= S(\hat{\rho}^{diag})- S(\hat{\rho}),
\label{eq-E-C}
\end{equation}
where the relative entropy of coherence $C(\hat{\rho})$ \cite{baumgratz} (from now on denoted as coherence/global coherence) will be employed as an appropriate measure of the system coherence, always with respect to the Hamiltonian \eqref{eq-Ham-N} eigenbasis. The definition makes use of the relative entropy $S(\hat{\rho}||\hat{\sigma})= {\rm Tr}(\hat{\rho}\ln\hat{\rho})-{\rm Tr}(\hat{\rho}\ln\hat{\sigma})$ and $\hat{\rho}^{diag}$ is the diagonal part of $\hat{\rho}$ \cite{baumgratz}.

The choice of the relative entropy of coherence, Eq.~\eqref{eq-E-C}, as our measure is motivated by its recognized connection with thermodynamic quantities \cite{kammerlander} through  the notion of von Neumann entropy.

Despite the fact the measure $C(\hat{\rho})$ is well established, it has one drawback, that can be noted in \cite{Gumberidze2019}, already. For multipartite system state $\hat{\rho}$, $C(\hat{\rho})$ is unable to distinguish the coherence contributions coming from the {\it local} states of each TLS and the coherence of the global state $\hat{\rho}$. This is a pure observation, not a critique, as this was not the purpose for which $C(\hat{\rho})$ was established. In general, such drawback can be noted, e.g., in certain optimization schemes in which the optimization with respect to the coherence $C(\hat{\rho})$ is carried out, while it may lead to a solution increasing the coherence {\it only locally}. We explicitly state, that this is not the case throughout this paper, as the pairwise protocol used here relies on the {\it global} structure of the measurement operators applied. We illustrate this fact by examining the properties of quantity $C_m(\hat{\rho})$, which we call {\it mutual relative entropy of coherence} (mutual coherence, for short), defined as \cite{XiSciRep2015,GuoPhysRevA2017,WangSciRep2017,TanPhysRevLett2018,Kraft_2018,YaoPRA2015,MaPRL2016}, 
\begin{equation}
C_{m}(\hat{\rho})= S (\hat{\rho}||\bigotimes_{i}\hat{\rho}_{i}) - S(\hat{\rho}^{diag}||\bigotimes_{i}\hat{\rho}_{i}^{diag})
\label{eq-C-mut-gen}
\end{equation}
where $\hat{\rho}$, $\hat{\rho}_{i}={\rm Tr}_{\forall j\neq i}(\hat{\rho})$ denote the total state of the system and the local states of its constituents, respectively, whereas $\hat{\rho}^{diag}$, $\hat{\rho}_{i}^{diag}$ are the diagonal (or fully dephased) versions of states. The mutual coherence thus specifies how much the distance (characterized by the relative entropy) of the state $\hat{\rho}$  with respect to the corresponding product of local states $\bigotimes_{i=1}^{N}\hat{\rho}_{i}$ increases compared to distance between diagonal (fully dephased) versions of that states. The RHS of Eq.~\eqref{eq-C-mut-gen} can be, after some algebra, simplified to 
\begin{equation}
C_{m}(\hat{\rho})= C(\hat{\rho})- C^{\it loc}(\hat{\rho}),
\label{eq-C-mut}
\end{equation}
where $C^{\it loc}(\hat{\rho})\equiv\sum_i C(\hat{\rho}_i)$, with the local states $\hat{\rho_i}$ defined as in \eqref{eq-C-mut-gen}. {\color{black} The mutual coherence, Eq.~\eqref{eq-C-mut}, quantifies as well the difference in work extracted \cite{kammerlander} within certain thermodynamic process from a coherent state $\hat{\rho}$, if the extraction is performed globally (with the total state) or locally (using the marginal states only).} Such quantity is, by definition, insensitive to the state transformations which increase only the local coherence. This is particularly suitable for our purpose of synthesizing and charging jointly large (global) coherent system from small (local), almost discharged cells \cite{Gumberidze2019}. 

\subsection{\label{pairwise_protocol} Pairwise protocol}
First, we consider $N$ copies of TLS in pure states, each characterized by the Hamiltonian \eqref{eq-Hams}
\begin{equation}
\ket{\psi_j}=\sqrt{p}\;\ket{e_j}+\sqrt{1-p}\;\ket{g_j},\quad j=1,\dots ,N.
\label{eq-psi-pure}
\end{equation}
The corresponding initial state of the total system reads
\begin{equation}
\ket{\Psi_{i}^{(N)}}=\displaystyle{\bigotimes_{j=1}^{N}\ket{\psi_j}},
\label{eq-pure-Psi-ini-N}
\end{equation}
with the initial energy determined with respect to the Hamiltonian \eqref{eq-Ham-N}
\begin{equation}
E_{0}^{(N)}=N E_{0}=\frac{NE}{2}(2p-1),
\label{eq-pure-E-ini-N}
\end{equation}
where $E_0$ is the energy of a single TLS, and the initial coherence $C_0^{(N)}=\sum_{j=1}^NC(\ket{\psi_j}\bra{\psi_j})$
\begin{equation}
C_0^{(N)}=-\sum_{k=0}^{N}\binom{N}{k}p^{N-k}(1-p)^k\ln\left[{p^{N-k}(1-p)^k}\right],
\label{eq-pure-C-ini-N}
\end{equation}
for more details see \cite{Gumberidze2019}.

The same system of TLS, characterized by the same initial state \eqref{eq-pure-Psi-ini-N}, was considered in our previous work \cite{Gumberidze2019}, where an alternative {\it global} charging protocol with a single measurement on $N$ TLS was employed. However, such protocol might be challenging from the point of view of a potential experimental realization, due to the high dimension and necessity of projecting on global pure state. Therefore, we examine in the following an alternative approach that could be experimentally less challenging, as it is based only on pairwise (of dimension four) projections, acting sequentially on the {\it pairs} of TLS. Moreover, the ground state elimination by pairwise projectors proved its universality in increasing the energy and coherence, jointly, while being diagonal in the energy basis \cite{Gumberidze2019}.

Namely, we employ pairs of orthogonal projectors, $\{\op{P}_0^{(j,k)},\; \op{P}_1^{(j,k)} \}$
\begin{equation}
\op{P}_0^{(j,k)}=\displaystyle{\ket{g_jg_k}\bra{g_jg_k}},\quad  \op{P}_1^{(j,k)}=\op{1}-\op{P}_0^{(j,k)},
\label{eq-proj-pair}
\end{equation}
where subscripts ``0'' and ``1'' stand for failure and success, respectively. These projectors act on a pair of $j$-th and $k$-th TLS.

In principle, $N$-TLS can be paired in an arbitrary fashion in the process of measurement. In this work we consider the general case of measuring TLS in successive pair sequence, i.e., the first and the second TLS, then the second and the third TLS, etc, see Fig.~\ref{scheme-2}.

In any scenario, the total projector on $N$-TLS, $\op{P}_1^{(N)}$, is invariant with respect to completely successful measurement sequence and constitutes the product of local projectors $\op{P}_1^{(j,k)}$
\begin{equation}
\op{P}_1^{(N)}=\displaystyle{\prod_{\substack{j=1 \\ k=j+1}}^{N} \op{P}_1^{(j,k)},\quad \op{P}_0^{(N)}=\op{1}-\op{P}_1^{(N)}}.
\label{eq-proj-pair-total}
\end{equation}

In other words, the order in which we measure the pairs of TLS does not play any role as in the end it will not affect the final state of the system, as the different projectors \eqref{eq-proj-pair} commute. Application of the projector $\op{P}_1^{(N)}$ to the state \eqref{eq-pure-Psi-ini-N} results in the successfully charged state 
\begin{equation}
\ket{\Psi_{f}^{(N)}}=\frac{\op{P}_1^{(N)}\ket{\Psi_i^{(N)}}}{\sqrt{p_s^{(N)}}},
\label{eq-pure-Psi-fin-N}
\end{equation}
with the following probability of success $p_s^{(N)}$, corresponding final energy $E_{f}^{(N)}$, and final coherence $C_{f}^{(N)}$ of the state, respectively
\begin{widetext}
\begin{equation}
p_s^{(N)}=\sum_{k=0}^{1}\binom{N}{k} \;p^{N-k} \; (1-p)^{k}+\sum_{k=2}^{t} \binom{N-k+1}{k} \; p^{N-k} \;(1-p)^{k}\neq 0,
\label{eq-pure-succ-N}
\end{equation}
\begin{equation}
E_{f}^{(N)}=\frac{E}{2 \; p_s^{(N)} } \left[\sum_{k=0}^{1} \; \binom{N}{k} \; (N-2k) \; p^{N-k}\;(1-p)^{k}+\sum_{k=2}^{t} \; \binom{N-k+1}{k} \; (N-2k) \; p^{N-k}\; (1-p)^{k} \right],
\label{eq-pure-Ef-N}
\end{equation}
\begin{equation}
C_{f}^{(N)}=-\sum_{k=0}^{1} \binom{N}{k} \;\frac{p^{N-k}\; (1-p)^{k}}{p_s^{(N)}}  \ln\left[\frac{p^{N-k}\; (1-p)^{k}}{p_s^{(N)}} \right]-\sum_{k=2}^{t} \binom{N-k+1}{k} \; \frac{p^{N-k}\; (1-p)^{k}}{p_s^{(N)}} \ln\left[\frac{p^{N-k}\; (1-p)^{k}}{p_s^{(N)}} \right],
\label{eq-pure-Cf-N}
\end{equation}
\end{widetext}
where $t=(N+1)/2$ if the number of TLS $N$ is odd and $t=N/2$ if $N$ is even. Due to the complex form of the results \eqref{eq-pure-succ-N}-\eqref{eq-pure-Cf-N} valid for the pure initial states \eqref{eq-pure-Psi-ini-N} only, we will present an approximate version of these quantities suitable for $p\ll 1$ region, see sec.~\ref{approximations}, and provide graphical representation of the results for mixed states dephased due to the presence of environment, see sec.~\ref{dephasing}.

In the following, we will be interested in whether the protocol \eqref{eq-proj-pair-total} increases the energy \eqref{eq-pure-DE-N}, $\Delta E^{(N)}>0$, as well as the coherence \eqref{eq-pure-DC-N}, $\Delta C^{(N)}>0$. The energy of the final successful state yields the corresponding increase in energy, see Eqs.~\eqref{eq-pure-E-ini-N}, \eqref{eq-pure-Ef-N}
\begin{equation}
\Delta E^{(N)}\equiv \frac{E_{f}^{(N)}-E_{0}^{(N)}}{E},
\label{eq-pure-DE-N}
\end{equation}
normalised by energy gap $E$, see Fig.~\ref{fig-delta-E} for an example.

The expression for the final coherence $C_f^{(N)}$, \eqref{eq-pure-Cf-N}, of the successful measurement outcome (the term corresponding to the von Neumann entropy vanishes for pure initial states) should be compared to the coherence of the initial pure state, Eq.~\eqref{eq-pure-C-ini-N}, yielding the definition of the coherence increase
\begin{equation}
\Delta C^{(N)}\equiv C_{f}^{(N)}-C_{0}^{(N)},
\label{eq-pure-DC-N}
\end{equation}
whereas the complete expression is omitted here for the sake of simplicity.

Based on observation made by comparison of Fig.~\ref{fig-delta-C}-\ref{fig-delta-Cm}, we point out one more positive aspect of the proposed protocol. The change of mutual coherence gives an additional information about the protocol's effect on the system coherence. If we compare Fig.~\ref{fig-delta-C} to Fig.~\ref{fig-delta-Cm} we can notice that for larger numbers $N$ of TLS the gain of mutual coherence is larger than corresponding coherence gain. This suggests that the part of global coherence of the system of $N$ TLS is increased. Therefore, not only the charging protocol increases the coherence of the total system, but it also transforms (consumes) the initial coherence (which equals to the sum of local coherences) into qualitatively different global final coherence. More formally, using the definitions in Eqs.~\eqref{eq-C-mut}, \eqref{eq-pure-DC-N} and labeling the local coherence according to Eq.~\eqref{eq-C-mut} by ``${(N)loc}$'' superscript and all coherences by subscripts ``$f$'' and ``$0$'' for final and initial state, respectively, we observe
\begin{eqnarray}
\nonumber
\Delta C_m^{(N)} &>&\Delta C^{(N)}\\
\nonumber
C_f^{(N)}-C_f^{(N)loc} &>& C_f^{(N)}-C_0^{(N)}=C_f^{(N)}-C_0^{(N)loc}\\
C_f^{(N)loc}&<& C_0^{(N)loc}, 
\label{eq-loc-coh-increase}
\end{eqnarray}
where we have used the fact that initial state \eqref{eq-pure-Psi-ini-N} is a product state, hence $C_0^{(N)}=C_0^{(N)loc}$. 

As another point, we may note the reversal of the curves' colors in comparing Figs.~\ref{fig-delta-C} and \ref{fig-delta-Cm}. This is caused by their opposite (decreasing vs. increasing) type of $p$ dependence for even $N$, see app.~\ref{pure_states_append}.

In the following paragraph, we try to explain the essence of the protocol's working principle. It relies on the fact that for low excitation probability $p$ ($p\ll 1$) of the initial state \eqref{eq-psi-pure} of each TLS, the global state of $N$ TLS, Eq.~\eqref{eq-pure-Psi-ini-N}, has significant occupation {\it only} in the lowest-lying energy eigenspaces close to the ground state of the total Hamiltonian \eqref{eq-Ham-N}. This occupation typically decreases rapidly (for low $p$) with the subspace energy eigenvalue, implying low values of the initial state energy $E_{0}^{(N)}$ and coherence $C_{0}^{(N)}$. Application of the pairwise protocol completely eliminates occupation of the energy eigenspaces in approximately the lower half of the global energy spectrum of the Hamiltonian \eqref{eq-Ham-N}. On contrary, it proportionally increases several occupations of the upper half of the spectrum, keeping the higher-energy level occupations flat distributed with different values in respective energy subspaces. Such redistribution of occupation always (for any $p$) increases the total average energy. For $p\ll 1$ the protocol increases as well  the final state coherence $C_{f}^{(N)}$ with respect to the initial state, because the initial populations are close to the ground state. For higher $p$ ($p\approx 1/2$) the initial state has populations similar to the (maximally coherent) flat distribution over all the energy eigenspaces. The protocol-induced wipe out of the low-energy populations drives the final state away from the initial high-coherence state (typically to the state that is flat distributed {\it only} in subspaces of the upper half of the spectrum), causing overall decrease of the final coherence $C_{f}^{(N)}$ relative to the initial one. Such qualitative picture holds in the case of even number $N$ of TLS entering the protocol. If, on contrary, $N$ is odd and $p\ll 1$, the protocol does wipe out all but a single mid-energy level completely, being responsible for lower $\Delta E^{(N)}$ and $\Delta C^{(N)}\rightarrow 0$ gains, cf. Fig.~\ref{fig-delta-E}-\ref{fig-delta-C}.

To enhance the chance to obtain the energy and coherence increase described above, we can employ the repeat until success (RUS) strategy \cite{Gumberidze2019}. Such strategy decreases the probability of failure $p_f=(1-p_s)^R$, with $p_s$ given in Eq.~\eqref{eq-pure-succ-N}. It relies on the possibility to recycle the TLS that fail to yield the successful outcome after the projector from Eq.~\eqref{eq-proj-pair} is applied. In such case all the TLS used for the synthesis can be sent back to the bath and the charging protocol can be reinitialized, see Fig.~\ref{scheme-2}, step 2). With increasing number of repetitions $R$, the total probability of successfully obtaining the final state \eqref{eq-pure-Psi-fin-N} approaches one, cf. Fig.~\ref{fig-prob-of-fail}.

As one can anticipate, creation of exact copies of TLS might be experimentally challenging, therefore we have considered and numerically checked stability of the out-coming $\Delta E^{(N)}>0$ and $\Delta C^{(N)}>0$ for a set of $N$ TLS with random initial $p$ sampled from a flat-distributed $p$-values of the width up to $\Delta p \approx 0.05$. As far as the probabilities of excitation differ from each other by less than few percent, the protocol is still applicable. For differences of the order $\Delta p \gtrsim 0.1$, we have observed detrimental effect on the coherence gain, while the energy gain was still present.   

\subsection{\label{approximations} Approximations}

In this subsection, we give approximate results describing the behavior of the quantities of interest, $\Delta E^{(N)} > 0$, $\Delta C^{(N)} > 0$, $\Delta C_m^{(N)}> 0$, matching the exact numerical results well in the domain $N\lesssim 10$, $p\lesssim 0.1$, if not specified otherwise. Such region of probability values $p$ has been chosen as it satisfies $\Delta E^{(N)} > 0$ jointly with $\Delta C^{(N)} > 0$ for the initial state \eqref{eq-pure-Psi-ini-N}, see Figs.~\ref{fig-delta-E-p}-\ref{fig-delta-C-p}. Based on the observation made from Fig.~\ref{fig-delta-C}-\ref{fig-delta-Cm}, we explicitly distinguish between odd and even number $N$ of TLS. For the purposes of our protocol it is preferable to employ {\it even} $N$, as it guaranties reaching better outcomes (see the qualitative explanation at the end of previous subsection), cf. App.~\ref{append_pairwise}.  

We start with the form of $p_s^{(N)}$, Eq.~\eqref{eq-pure-succ-N}. Its behavior corresponds to the power functions 
\begin{eqnarray}
\nonumber
p_s^{(N)} &\approx & p^{\frac{N-1}{2}},\; {\rm odd}\, N\\ 
p_s^{(N)} &\approx & \left(\frac{N}{2}+1\right) p^{\frac{N}{2}},\; {\rm even}\, N
\label{eq-approx-ps}
\end{eqnarray}
in the range of small probability of excitation $p\ll 1$. For even $N$ the accuracy of this approximation is kept within $10 \%$.

The energy gain, $\Delta{E}^{(N)}$, for odd $N$  depends linearly on $N/2$, the number of TLS in a good approximation up to $p^{0}$, whereas for even number of TLS the accuracy of the approximation improves significantly by including the linear term in $p$, resulting to 
\begin{eqnarray}
\nonumber
\Delta{E}^{(N)}&\approx  &\frac{N-1}{2},\;{\rm odd}\, N\\
\Delta{E}^{(N)}&\approx &\frac{N}{2}-\frac{N(20-N)}{24}p,\;{\rm even}\, N.
\label{eq-approx-DE}
\end{eqnarray}
For even $N$ approximation, see the full line in Fig.~\ref{fig-delta-E}.

As the considerable increase in coherence $\Delta C^{(N)}>0$ is observed only in  region of relatively small $p$, cf. Fig.~\ref{fig-delta-C-p}, we can find useful approximation giving us the general behavior in this region of $p$, yielding
\begin{eqnarray}
\label{eq-pure-DC-approx}
\Delta C^{(N)}&\approx &\frac{(N-1)(N-3)}{8}(1-\ln{p})p,\,{\rm odd}\, N\\
\nonumber
\Delta C^{(N)}&\approx &\ln{\left(\frac{N}{2}+1\right)}-\frac{N(20-N)}{24}(1-\ln{p})p,\,{\rm even}\, N.
\end{eqnarray}
The even $N$ approximations are shown as full lines in Fig.~\ref{fig-delta-C}, or as dashed lines in Fig.~\ref{fig-delta-C-p}.

It should be noted, that the above approximations do not match the exact numerical results (discrete points in Fig.~\ref{fig-delta-E}-\ref{fig-delta-Cm}, dashed lines in Figs.~\ref{fig-delta-E-p}-\ref{fig-delta-Cm-p}) equally well. Much tighter correspondence, using these simplest approximations, is found for $N$ even. This is convenient, as this subset works better when used in the protocol, although its accuracy is lost with the increased number $N$ of TLS in all cases.

For the same reasons (better match and higher values) we list here the even $N\lesssim 10$, small $p$ approximation of the mutual coherence $C_{m}$, defined in Eq.~\eqref{eq-C-mut}, here applied to the  final state \eqref{eq-pure-Psi-fin-N} of our protocol, here labeled as $\Delta C^{(N)}_{m}$. Hence, for even $N$ it can be well approximated  in the region of $N\lesssim 10$, $p\leqslant 0.12$ (with small relative deviation $\lesssim 2 \%$) as
\begin{eqnarray}
\nonumber
\Delta C^{(N)}_{m}&\equiv & C^{(N)}_{m,f} = C^{(N)}_{f} - C^{(N)loc}_{f}\\  
\nonumber
&\approx &\ln{\left(\frac{N}{2}+1\right)}+ \frac{N(N+4)}{24}(1-\ln{p})p\\
&&-\ln{\left[\frac{3}{2}\left(\frac{N}{2}+1\right)!\right]}p,\;{\rm even}\, N.
\label{eq-pure-DCm-approx}
\end{eqnarray}
The last term in the RHS approximates the local coherence of the system after the measurement, $C^{(N)loc}_{f}$.

The suitability of the above-mentioned approximations for $p\ll 1$ is shown in Figs.~\ref{fig-delta-E}-\ref{fig-delta-Cm} and Fig.~\ref{fig-prob-of-succ}. It can be seen that exact values, represented by dots, are well fitted by respective approximate expressions, represented by solid lines. Comparison of RHS of Eqs.~\eqref{eq-pure-DC-approx} and \eqref{eq-pure-DCm-approx} also reveals the reason for reversal of the curves' colors in Figs.~\ref{fig-delta-C} and \ref{fig-delta-Cm}. It is due to the opposite sign of the $p$-dependence of these approximations. The same reason is responsible for the color reversal also in the models including dephasing, see Sec.~\ref{dephasing}.

{\color{black} We can compare results of our protocol to the results of {\it optimal} probabilistic distillation of pure states towards maximally coherent state in a given dimension, see e.g. \cite{adessoPRL2018}. Taking into account that our initial state is fixed, Eq.~\eqref{eq-pure-Psi-ini-N}, and the final target state would be  $\ket{\Psi_m}=1/\sqrt{m}\sum_{i=1}^m\ket{i}$ with $m=2^N$, such protocol optimized for global coherence gain \eqref{eq-pure-DC-N} with use of strictly incoherent \cite{adessoPRL2018} operations, yields for $p\ll 1$
\begin{eqnarray}
\Delta C^{(N)}_{\rm opt}=N\left [\ln{2}-p(1-\ln{p})\right ],
\label{eq-pure-DC-optimal}
\end{eqnarray}
being clearly superior (by construction $\propto N$) to our results  \eqref{eq-pure-DC-approx} for even $N$ ($\propto\ln N$). However, such protocol will certainly loose the universality feature, as the used strictly incoherent operation would necessarily be input state-dependent.
On contrary, the success probability of reaching the maximally coherent state  \cite{adessoPRL2018}, again using the input state \eqref{eq-pure-Psi-ini-N}, reads
\begin{eqnarray}
p_{s,{\rm opt}}^{(N)}=(2p)^N,
\label{eq-pure-ps-optimal}
\end{eqnarray}
being always smaller (for $p\ll 1$) than the even $N$ case of \eqref{eq-approx-ps}, showing an existing trade-off between attainable final state coherence and the success probability of achieving it. Simple discussion of mutual coherence increase of our protocol vs. the above mentioned global coherence-gain optimized one can not be performed. The class of maximally coherent states in Hilbert space with dimension $m=2^N$ composed of tensor product of respective subsystems' Hilbert spaces includes separable, as well as maximally entangled states. Therefore, the mutual coherence values can span the whole interval $[0,N\ln 2]$. For this reason, such discussion exceeds the scope of this paper.}

We list here some results of our previous work \cite{Gumberidze2019} as well, where we have used a single {\it global} measurement on all TLS instead of the pairwise protocol presented here. The approximate results obtained from global method are plotted as dashed, continuous lines in Figs.~\ref{fig-delta-E}-\ref{fig-delta-C} and Figs.~\ref{fig-prob-of-succ}-\ref{fig-prob-of-fail}. The respective approximations in the case of the global protocol are
\begin{eqnarray}
\nonumber
p_s^{(N)} &\approx & Np, \\
\nonumber
\Delta{E}^{(N)} &\approx & 1-\frac{N+1}{2}p,\\ 
\nonumber
\Delta{C}^{(N)} &\approx & \ln{N}-\frac{N+1}{2}(1-\ln{p})p,\\
\nonumber
\Delta{C}_{m}^{(N)} &\approx & \ln{N}+\frac{N-1}{2}(1-\ln{p})p\\
& - & \frac{(N-1)^{2}}{N-2}\ln{(N-1)}p.
\label{eq-global-protocol-approx}
\end{eqnarray}
As we have checked numerically, the relative deviation $\lesssim 10\%$ of the global approximations holds in the range $p \lesssim 0.02$ for $p_s^{(N)}$, $p \lesssim 0.08$ for $\Delta{E}^{(N)}$, $p \lesssim 0.06$ for $\Delta{C}^{(N)}$, and $p \lesssim 0.11$ for $\Delta{C}_m^{(N)}$ (for all $N\lesssim 8$).

Looking at the global results, Eqs.~\eqref{eq-global-protocol-approx}, we can qualitatively compare the results for both protocols within the $p\ll 1$ range. As for the $p_s^{(N)}$, the results of the global protocol  increase linearly with $N$, while pairwise results decrease exponentially. Thus, the global protocol has a noticeable advantage over the pairwise case, in this respect. The energy gain, $\Delta{E}^{(N)}$, decreases with $N$ in the global case, compared to an increase with $N$ in the pairwise approach. For the coherence gain, $\Delta{C}^{(N)}$, the qualitative behavior is the same, whereas the increase is lower for the pairwise protocol, see Fig.~\ref{fig-delta-E}-\ref{fig-delta-C}. The better performance of the global protocol in some of the previous points is, in our opinion, more than balanced by its expected increasing experimental complexity for larger values of $N$.

\section{\label{dephasing} COHERENCE SYNTHESIZED FROM mixed TLS states}
As any real physical system is interacting with its environment, the initial pure states of TLS might be negatively affected by such interaction. To consider the most basic decoherence mechanism, we assume partially and individually dephased TLS \cite{NielsenBook}, both before and after the synthetisation. Thus, the dephasing reduces the off-diagonal elements in the energy representation. Hence, of practical importance might be to explore the suitability of using these partially coherent states resulting from such dephasing process in our protocol. Thus, the effect of dephasing applied to the initial and final state of the system is considered below to test the charging protocol for stable performance.

\begin{figure}
\centering
\subfloat[\label{fig-delta-C-pre-9}]{
\includegraphics[width=0.95\columnwidth]{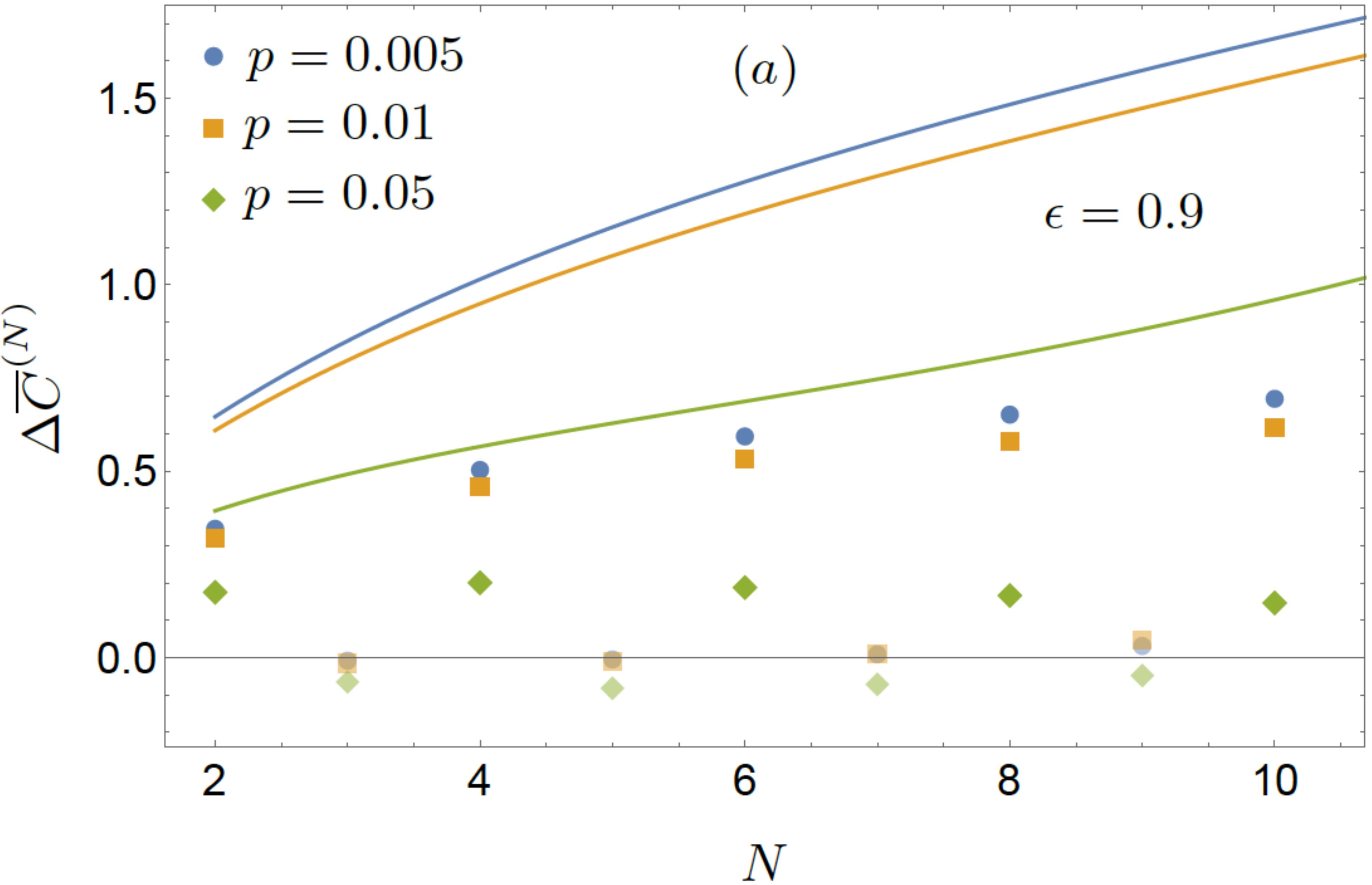}}
\hfill \vspace{-0.5cm}
\subfloat[\label{fig-delta-C-mut-pre-9}]{\includegraphics[width=0.95\columnwidth]{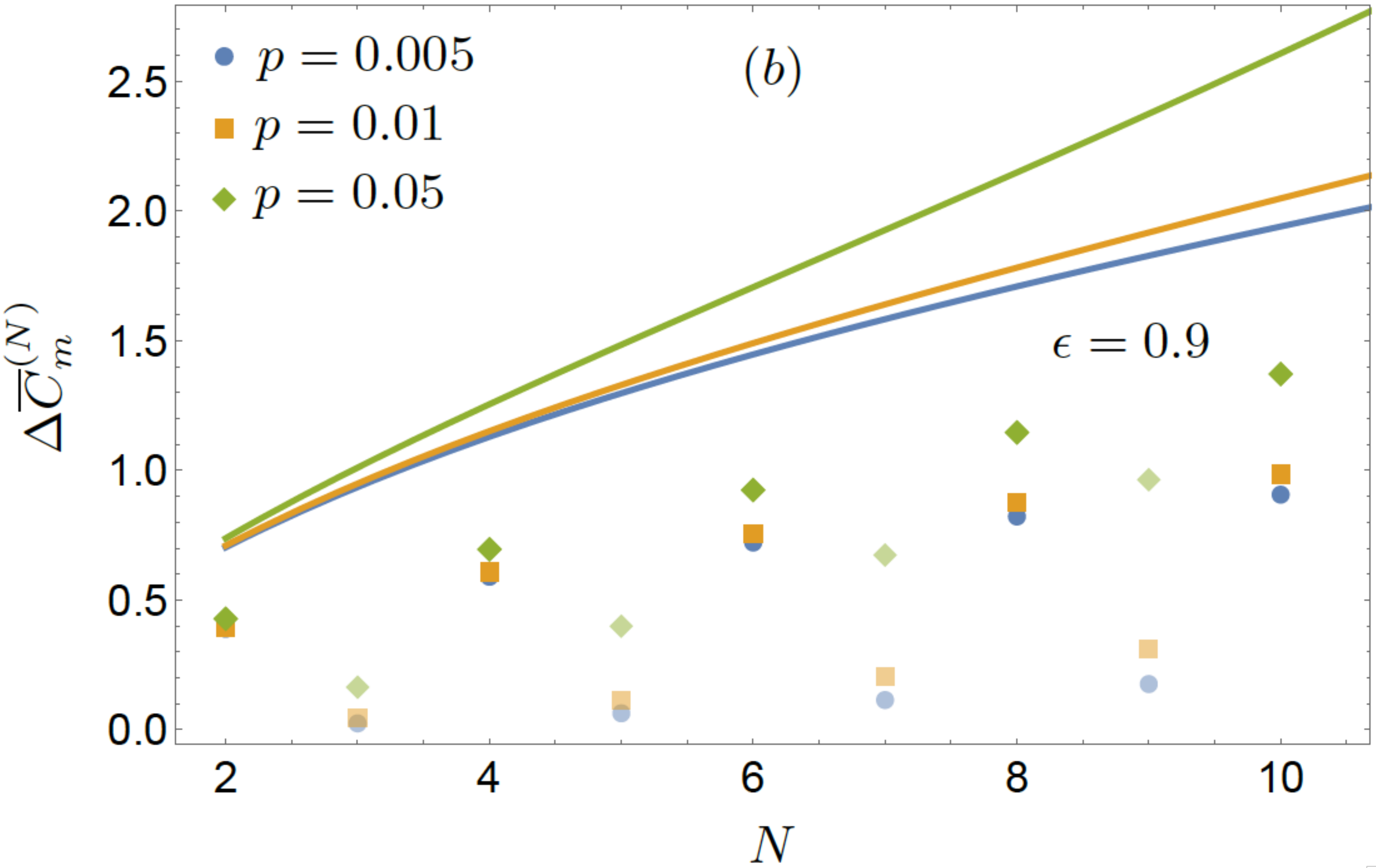}}
\vspace{-0.3cm}
\caption{Conditional synthesis of coherence and mutual coherence of $N$ partially dephased TLS. \protect\subref{fig-delta-C-pre-9} The coherence gain $\Delta \overline{C}^{(N)}$, Eq.~\eqref{eq-pure-DC-N}, and \protect\subref{fig-delta-C-mut-pre-9} the mutual coherence gain $\Delta \overline{C}_{m}^{(N)}$ for $N$ TLS in the final state \eqref{eq-mixed-fin} and dephased initial state \eqref{eq-mixed-N-tot} of TLS. The dephasing rate is set to $\epsilon=0.9$. The marked discrete values are results of exact numerical calculation. The full lines recall the approximate results for pure initial states, see Fig.~\ref{fig-delta-C}-\ref{fig-delta-Cm}.
The effect of the coherence gain decrease is significantly stronger for larger numbers of TLS constituting the system, causing the emergence of the local maximum at $N\approx 4$. Generally, the gains $\overline{C}^{(N)}>0$, $\overline{C}_{m}^{(N)}>0$ are preserved until the "critical" value $\epsilon\approx 0.5$, where it is lost even for small values of $p$ and low $N$. Note as well the same curves' color reversal as pointed out in Fig.~\ref{fig-delta-C}-\ref{fig-delta-Cm} and explained in Sec.~\ref{approximations}.}
\label{fig-delta-C-pre}
\end{figure}

\subsection{\label{sub-predephasing} Dephasing of the initial state of TLS}

In this subsection, we are going to explore the effect of dephasing of the initial states of TLS, Eq.~\eqref{eq-pure-Psi-ini-N}, to find out how it affects the results of the protocol. Let us assume that the initial pure state \eqref{eq-psi-pure} of each TLS has suffered from dephasing, thus is characterized by the density matrix \cite{NielsenBook,Gumberidze2019}
\begin{eqnarray}
\nonumber
\op{\rho}_j&=&p\ket{e_j}\bra{e_j}+\epsilon\sqrt{p(1-p)}(\ket{e_j}\bra{g_j}+\ket{g_j}\bra{e_j})\\
&&+(1-p)\ket{g_j}\bra{g_j},\quad j=1,\dots ,N,
\label{eq-mixed-N}
\end{eqnarray}
where $0\leq\epsilon\leq 1$ quantifies the effect of the dephasing.  The  initial state of the total system of $N$ TLS reads 
\begin{equation}
\hat{\rho}_{i}^{(N)} =\displaystyle{\bigotimes_{j=1}^{N}}
\hat{\rho}_j.
\label{eq-mixed-N-tot}
\end{equation}

For simplicity, we assume that all TLS undergo the dephasing process with the same value of $\epsilon$. As the energy is defined only by the diagonal terms of the density matrix, it is clearly not affected by the dephasing process. Hence $\Delta E^{(N)}$ does not change compared to the pure state case.  

On contrary, the coherence comprises \cite{baumgratz} the von Neumann entropy term having non-zero contribution in case of mixed states, which in turn is affecting the initial and final coherence, $\overline{C}_0^{(N)}$ and $\overline{C}_f^{(N)}$, resulting in the change of coherence difference, $\Delta \overline{C}^{(N)}$. The single bar labels quantities calculated with the effect of dephasing acting only on the initial TLS state.

Subjecting the initial state to the same projector-based charging procedure as in Sec.~\ref{pairwise_protocol}, 
$\{\op{P}_1^{(N)},\,\op{P}_0^{(N)}\}$, yields the final state
\begin{equation}
\label{eq-mixed-fin}
\op{\rho}_f^{(N)}=\frac{\op{P}_1^{(N)}\op{\rho}_i^{(N)}\op{P}_1^{(N)}}{{\overline{p}_s}}.
\end{equation}

The complexity of the coherence measure \eqref{eq-E-C} prevents us from determining $\Delta \overline{C}^{(N)}$ analytically for the initial state \eqref{eq-mixed-N-tot} and the final state \eqref{eq-mixed-fin}. Thus, we focus our attention to the fully numerical results, presenting them only graphically. 

These results for the dephased initial state of the system are shown in Fig.~\ref{fig-delta-C-pre-9} and Fig.~\ref{fig-delta-C-mut-pre-9} with the value $\epsilon=0.9$ for coherence and mutual coherence, respectively. It can be seen that dephasing of the initial state diminishes both coherence and mutual coherence gains, especially for even numbers $N$ of TLS. However, dephasing does not destroy the positive effect of the protocol completely. Quite generally, small dephasing, substantially below the "critical" value $\epsilon\approx 0.5$, may affect the system, while the coherence gain can still remain positive and non-negligible. The term "critical" is meant in the sense that such values of $\epsilon$ cause loss of the coherence gains, $\Delta \overline{C}^{(N)}\approx 0$ and $\Delta \overline{C}^{(N)}_m\approx 0$ for any $N$. For larger values of $\epsilon$, dephasing generally causes appearance of a local maximum for certain $N$, as with $N$ the negative effect of dephasing increases. We also point out that $\Delta \overline{C}^{(N)}$ and $\Delta \overline{C}^{(N)}_m$ keep irrespective of dephasing their decreasing and increasing dependence on $p$, respectively. This is manifested by the curves' opposite ordering with respect to values of $p$, as discussed in Sec.~\ref{approximations}, and the same reason holds for dephasing affecting the final  (post-protocol) state discussed in the next subsection.

In reality, each TLS's initial state may dephase in time with different rates $\epsilon_{1},\epsilon_{2},\dots,\epsilon_{N}$ as they are independent of each other, resulting in different total initial state of TLS before the measurement. We have considered such situation and  checked the protocol outcome numerically. For the sake of brevity, we present the results only briefly. The local dephasing with different values of $\epsilon$ for each TLS, has no truly detrimental effect. As long as the dephasing parameters are bigger than critical value $\epsilon \gtrsim 0.5$ and the respective values are close to each other $\epsilon_i\approx\epsilon_j$, the coherence and mutual coherence gains decrease, but qualitatively, the main effects survive.  



\begin{figure}
\centering
\subfloat[\label{fig-delta-C-pos-9}]{
\includegraphics[width=0.95\columnwidth]{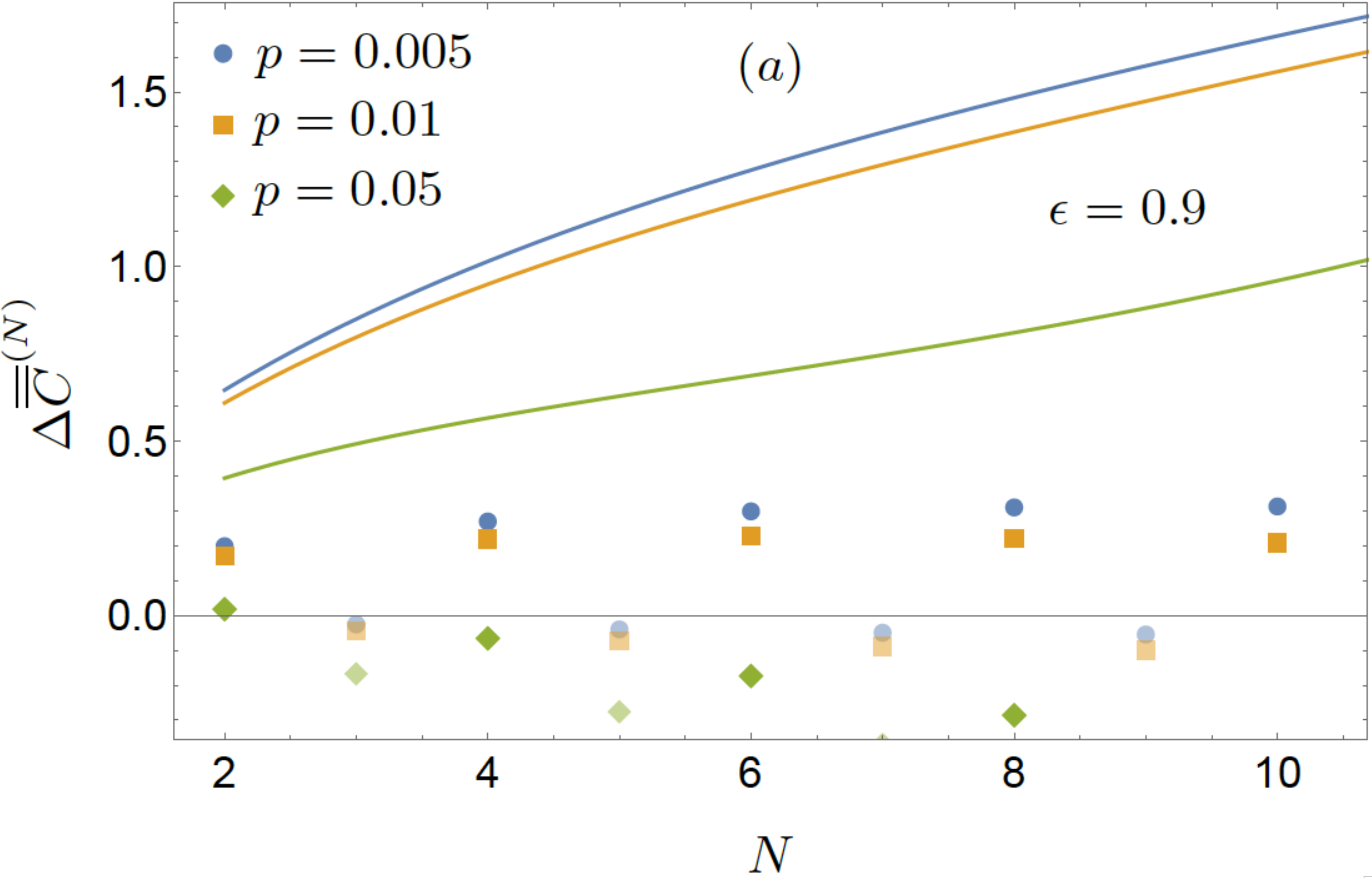}}
\hfill
\vspace{-0.5cm}
\subfloat[\label{fig-delta-C-mut-pos-9}]{\includegraphics[width=0.95\columnwidth]{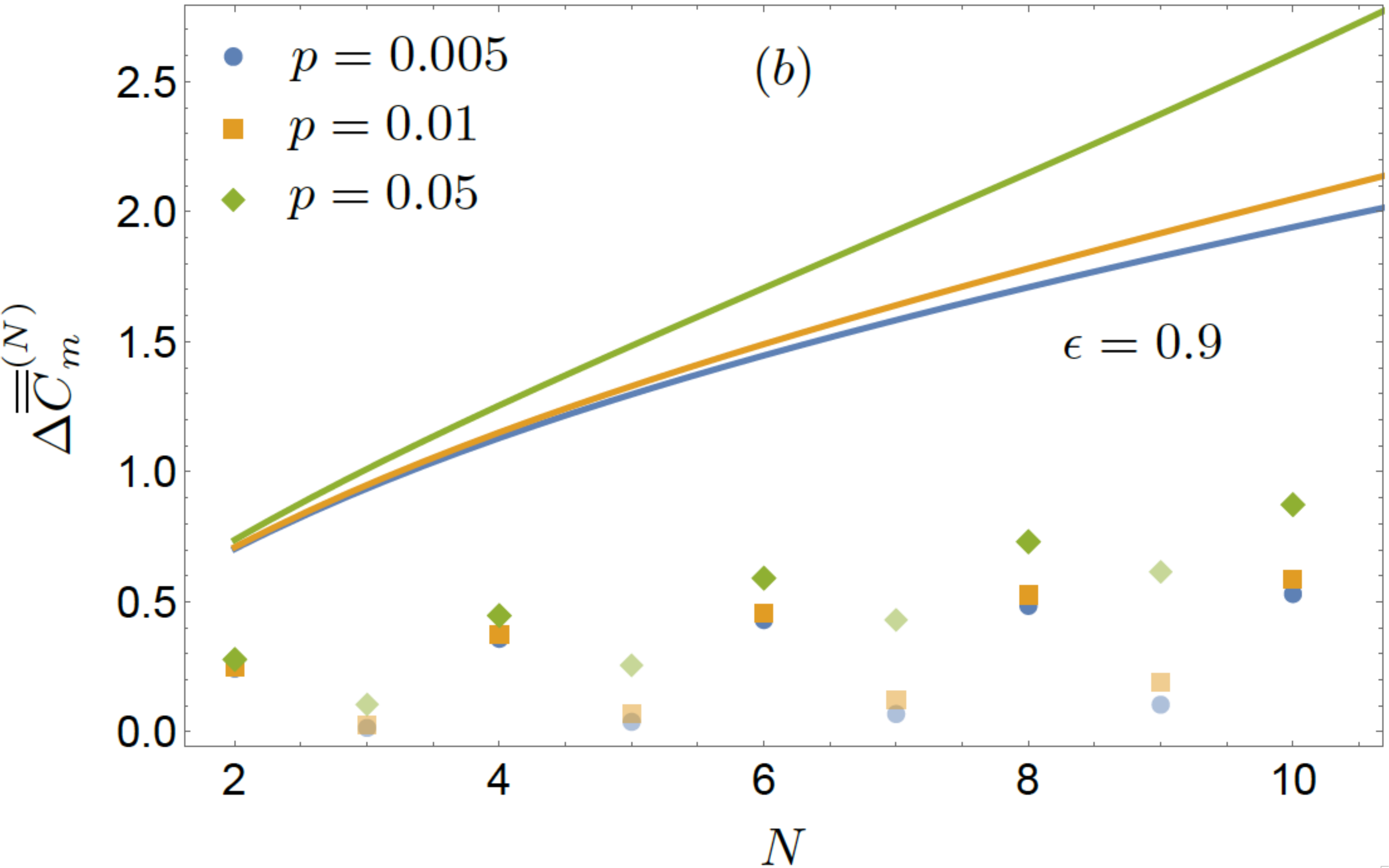}}
\vspace{-0.3cm}
\caption{Effect of additional dephasing on the final state \eqref{eq-mixed-fin}.  \protect\subref{fig-delta-C-pos-9} The coherence gain, $\Delta \overline{\overline{C}}^{(N)}$ and \protect\subref{fig-delta-C-mut-pos-9} the mutual coherence gain $\Delta \overline{\overline{C}}_{m}^{(N)}$  of the system in the final states, Eq.~\eqref{eq-mixed-fin-2}, with $\epsilon=0.9$. The values are plotted versus the number of TLS $N$, for different excitation probabilities $p$, Eq.~\eqref{eq-psi-pure}. Solid lines represent the corresponding values for pure states, cf. Fig.~\ref{fig-delta-C}-\ref{fig-delta-Cm}, marked discrete points are exact numerical results. It can be seen by comparison with Fig.~\ref{fig-delta-C-pre}, that dephasing of initial states has much larger effect of diminishing the coherence gains, than additional dephasing process of final state of the system. However, even in this case dephasing decreases the gain in $\Delta \overline{\overline{C}}^{(N)}$, for larger $N$ and low $p$, and causes the appearance of local maximum for $N\approx 6$ (orange squares).}
\end{figure}

\subsection{\label{postdephasing} Dephasing after the synthesization}
Different types of multipartite states have different sensitivity to (possibly local) influence of the environment, such as dephasing studied in the previous subsection. Hence, we are going to test the robustness of the system {\it final} state \eqref{eq-mixed-fin} with respect to the local effects of interaction with environment. Such dephasing of the system after the measurement results into
\begin{equation}
\label{eq-mixed-fin-2}
\overline{\op{\rho}}_f^{(N)}=\sum_{i=0}^{2^N-1}\op{\mathcal{K}}_i\op{\rho}_f^{(N)}\op{\mathcal{K}}_i^\dagger,
\end{equation}
with the final coherence $\overline{\overline{C}}_f^{(N)}\equiv C(\overline{\op{\rho}}_f^{(N)})$.  The $\op{\mathcal{K}}_i$ being the global Kraus operators \cite{kraus}, having the form
\begin{equation}
\op{\mathcal{K}}_i\equiv \op{K}_{j_{N-1}^{(i)}}\bigotimes\cdots\bigotimes\op{K}_{j_1^{(i)}}\bigotimes\op{K}_{j_0^{(i)}},\quad j_k^{(i)}=\{0,1\},
\label{eq-kraus-binary}
\end{equation}
with $(j_{N-1}^{(i)}\cdots j_1^{(i)}j_0^{(i)})_2=(i)_{10}$ being the binary representation of the index value $(i)_{10}$, e.g., $(7)_{10}=(0111)_{2}$ for $N=4$. The local Kraus operators transform the initial single qubit state $\op{\rho}_f^{(1)}$ into the final dephased state  $\op{\overline{\rho}}_f^{(1)}$ as
\begin{equation}
\op{\overline{\rho}}_f^{(1)}=\hat{K}_0\op{\rho}_f^{(1)}\hat{K}_0 + \hat{K}_1\op{\rho}_f^{(1)}\hat{K}_1,
\end{equation}
with the definitions
\begin{eqnarray}\nonumber
\op{K}_0=\sqrt{\frac{1+\epsilon}{2}} \hat{I},\quad \op{K}_1=\sqrt{\frac{1-\epsilon}{2}} \hat{\sigma}_z,
\end{eqnarray}
resulting into the form of Eq.~\eqref{eq-mixed-N}.
We will assume that TLS, even though they are in entangled state $\op{\rho}_f^{(N)}$~\eqref{eq-mixed-fin}, undergo the dephasing independently (locally), see structure of Eq.~\eqref{eq-kraus-binary}, with some constant value of the post-protocol $\overline{\epsilon}$, which can be generally different from dephasing parameter $\epsilon$ of the initial state $\op{\rho}_i^{(N)}$, Eq.~\eqref{eq-mixed-N}. However, for simplicity, in the following we will assume $\epsilon=\overline{\epsilon}$ being equal. {\color{black} Here, we want to stress out one important fact stemming from the diagonal form of the Kraus operators in Eq.~\eqref{eq-kraus-binary}. These diagonal operators commute with the, as well diagonal, projectors \eqref{eq-proj-pair} used in our protocol. This allows for possibility of formally interchanging the order of the second (post-protocol) dephasing and the measurements, in principle simplifying modelling of such environmental interaction. Although being aware of this fact, we retain the description of the second (post-protocol) dephasing in the form \eqref{eq-mixed-fin-2}, for conceptual and pedagogical reasons.}

It can be seen from Fig.~\ref{fig-delta-C-pos-9} and Fig.~\ref{fig-delta-C-mut-pos-9} for $\epsilon=0.9$, that the dephasing of the final state of the system affects the coherence gain by further decreasing it,  $\Delta \overline{\overline{C}}^{(N)} < \Delta \overline{C}^{(N)}$. But with the dephasing parameter $\epsilon$ bigger than "critical" value $0.75$, there still exists gain of the coherence, $\Delta \overline{\overline{C}}^{(N)} > 0$ in the range of small values of $p$. 

In the same way, the dephasing of initial and final states of the system affects the mutual coherence gain $\Delta \overline{\overline{C}}^{(N)}$, however the effect is weaker for larger $N$.

\section{\label{conclusions} Conclusions and outlook} 
We have introduced a feasible quantum coherence synthesizing protocol. It is based on sequential pairwise application of projective measurements on independent and non-interacting TLS with low initial coherence with respect to energy basis. These projectors are diagonal in the energy basis and conditionally remove ground states of the TLS pairs. The protocol synthesizes  small coherence of the local TLS states into a global coherence of the TLSs, scaling logarithmically with their number $N$, while jointly increasing their energy as well, that scales linearly with $N$, rendering the protocol universal from this perspective. The protocol is sufficiently robust with respect to the initial state dephasing induced by the environment, as well as to additional dephasing of the resulting state. 

The protocol universality, that ensures applicability for any weakly excited TLSs, simultaneously suggests its non-optimality in the sense that its results can be quantitatively improved in the desired direction. We are led to such expectation by the fact that as our protocol's resulting coherence scales with the number of TLS $N$ as $\propto\ln N$, the maximally coherent state of such system would have the coherence scaling $\propto N$, opening large space for improvement. Due to the problem dimension strongly increasing with the number of TLS, we expect any general optimization to become more complex, as well. Hence, we suggest as the possible improvement, increasing the number of universal projectors, e.g., by adding the projections of more distant TLS (last-first or other combinations), but still resorting to the pairwise-type operations. Such modifications may potentially increase the final values of the quantities, but presumably at the expense of the success probability. Another point which can be improved connects to the fact, that the current protocol works best in the low-excitation region, up to intermediate values, but breaks down in the region of highly initially excited TLS. In such case, the original protocol eliminating in each step the {\it ground state} populations of each measured pair, can be modified to eliminate the {\it excited states} populations, which can be anticipated based on symmetry arguments.

We believe that the protocol universality may be used as a spring board for others in the research directed toward quantum coherence manipulation \cite{HofheinzNat2009,LeghtasPhysRevA2013,SharmaPhysRevA2016,Wu17}, or exploited in the field of quantum thermodynamics, e.g., as a quantum battery charging protocol \cite{alicki2013, binderNJP2015,campaioliPRL2017}. 
The proof-of-principle test will require many TLS that can be coupled in a pairwise manner, projected, and their overall state evaluated by a quantum-state tomography. Therefore, long-standing trapped ions \cite{MonzPRL2011,FriisPRX2018}, progressing superconducting qubits \cite{Arute2019} and recent optical experiments \cite{WangPRL2019} can be used to test the predicted rules of quantum coherence synthesis.



\begin{acknowledgments}
M.G. acknowledges the support of the project No.~20-16577S of Grant Agency of the Czech Republic. R.F. and M.K. would like to acknowledge the support of project No.~19-19189S of the Grant Agency of the Czech Republic and the project IGA-PrF-2021-006.
\end{acknowledgments}


\appendix

\section{\label{append_pairwise} Pure states, dependence on p}
\label{pure_states_append}

\begin{figure}
\includegraphics[width=0.95\columnwidth]{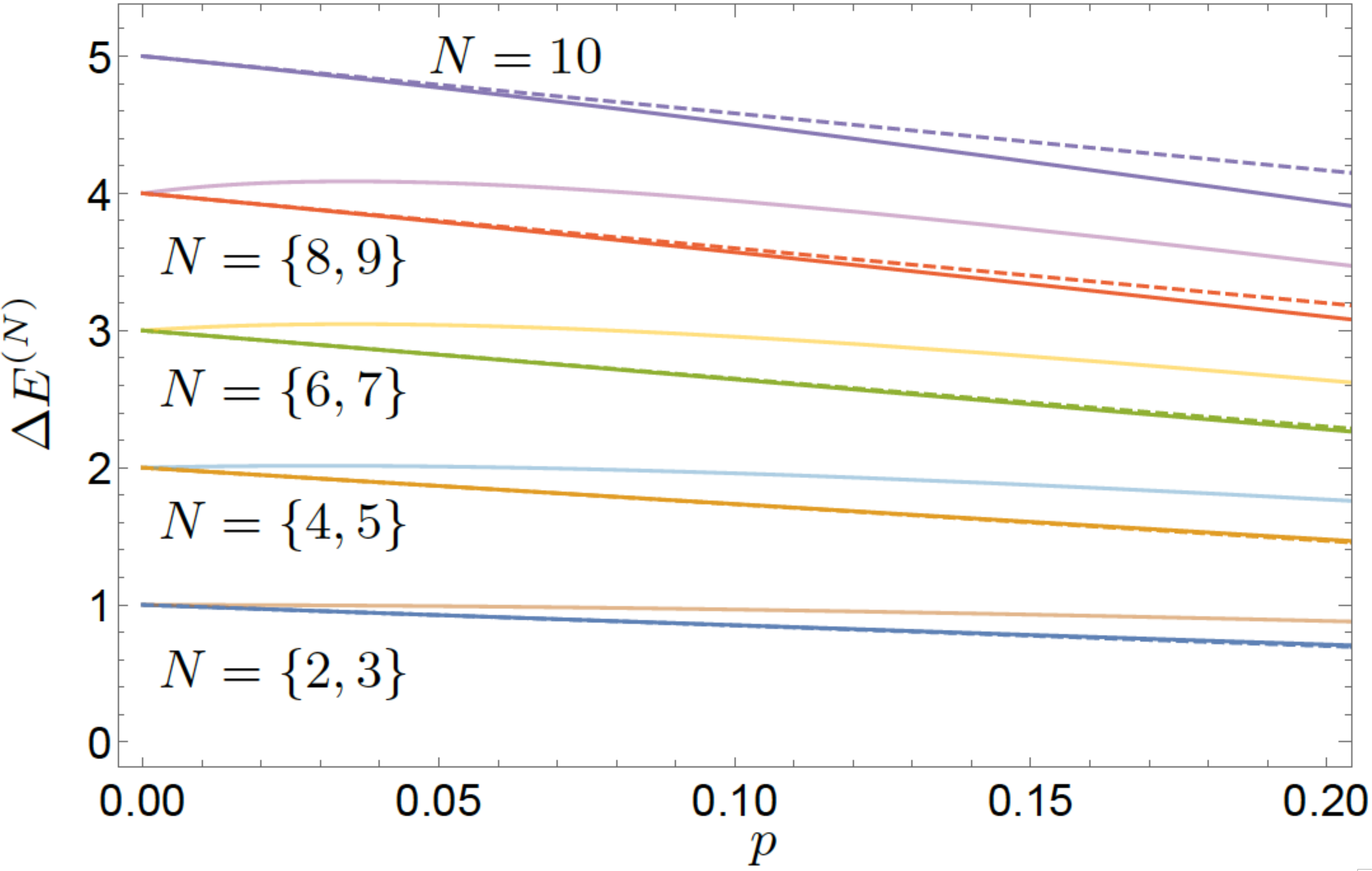}
\caption{The plot of the normalized average energy gain $\Delta E^{(N)}$, Eq.~\eqref{eq-pure-DE-N}. The values are plotted versus different excitation probabilities $p$, Eq.~\eqref{eq-psi-pure}, and parametrized by the number of TLS $N$. For odd $N$ the lines are made more transparent. Full lines correspond to exact results, the dashed lines show the corresponding approximations, Eq.~\eqref{eq-approx-DE}.
The energy gain $\Delta E^{(N)}>0$ increases proportionally to the number $N$ of TLS and decreases linearly with $p$ for even $N$, which we focus at, see Fig.~\ref{fig-delta-C-p} for the reason.}
\label{fig-delta-E-p}
\end{figure}

\begin{figure}
\includegraphics[width=0.95\columnwidth]{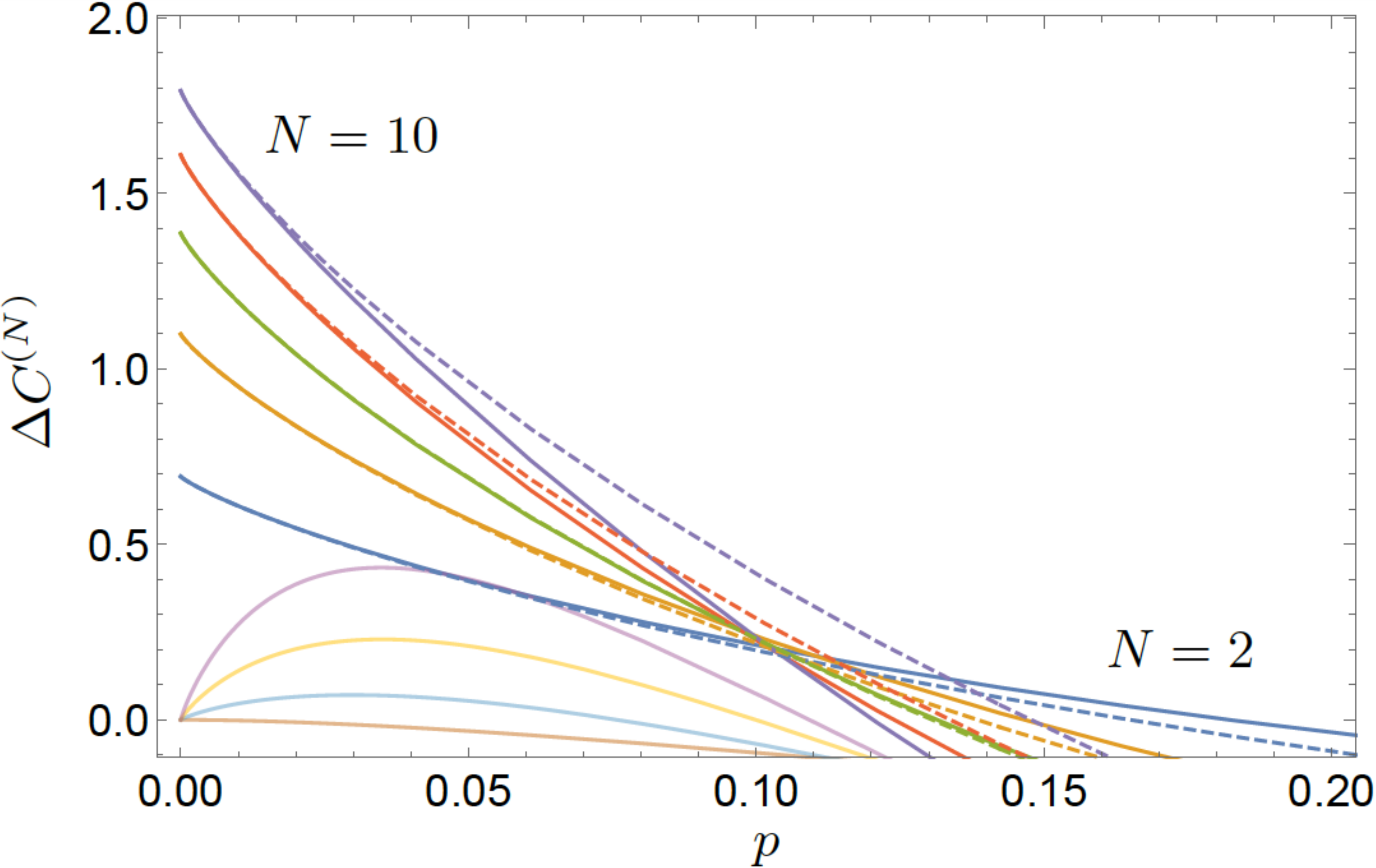}
\caption{Plot of the coherence gain $\Delta C^{(N)}$, Eq.~\eqref{eq-pure-DC-N},  dependence on the single TLS excitation probability $p$, Eq.~\eqref{eq-psi-pure}. The different curves are parametrized by different values of $N$, the more transparent ones are for odd $N$, revealing quantitatively lower values of $\Delta C^{(N)}$. The color code of the lines remains the same as in Fig.~\ref{fig-delta-E-p}. As seen from the plot, only low-excited TLS can be used to synthesize larger coherent system, as $\Delta C^{(N)}>0$ noticeably only in the $p\ll 1$ region. The dashed lines represent the corresponding approximations \eqref{eq-pure-DC-approx}. We stress again that the coherence gain  $\Delta C^{(N)}$ differs for odd and even $N$ with a significant advantage for even numbers of TLS used as an input for the protocol. Furthermore, we point out that the approximate results (dashed lines) are loosing their accuracy with increasing $N$ (even), being well valid in a shrinking interval of $p$.}
\label{fig-delta-C-p}
\end{figure}

\begin{figure}
\includegraphics[width=0.95\columnwidth]{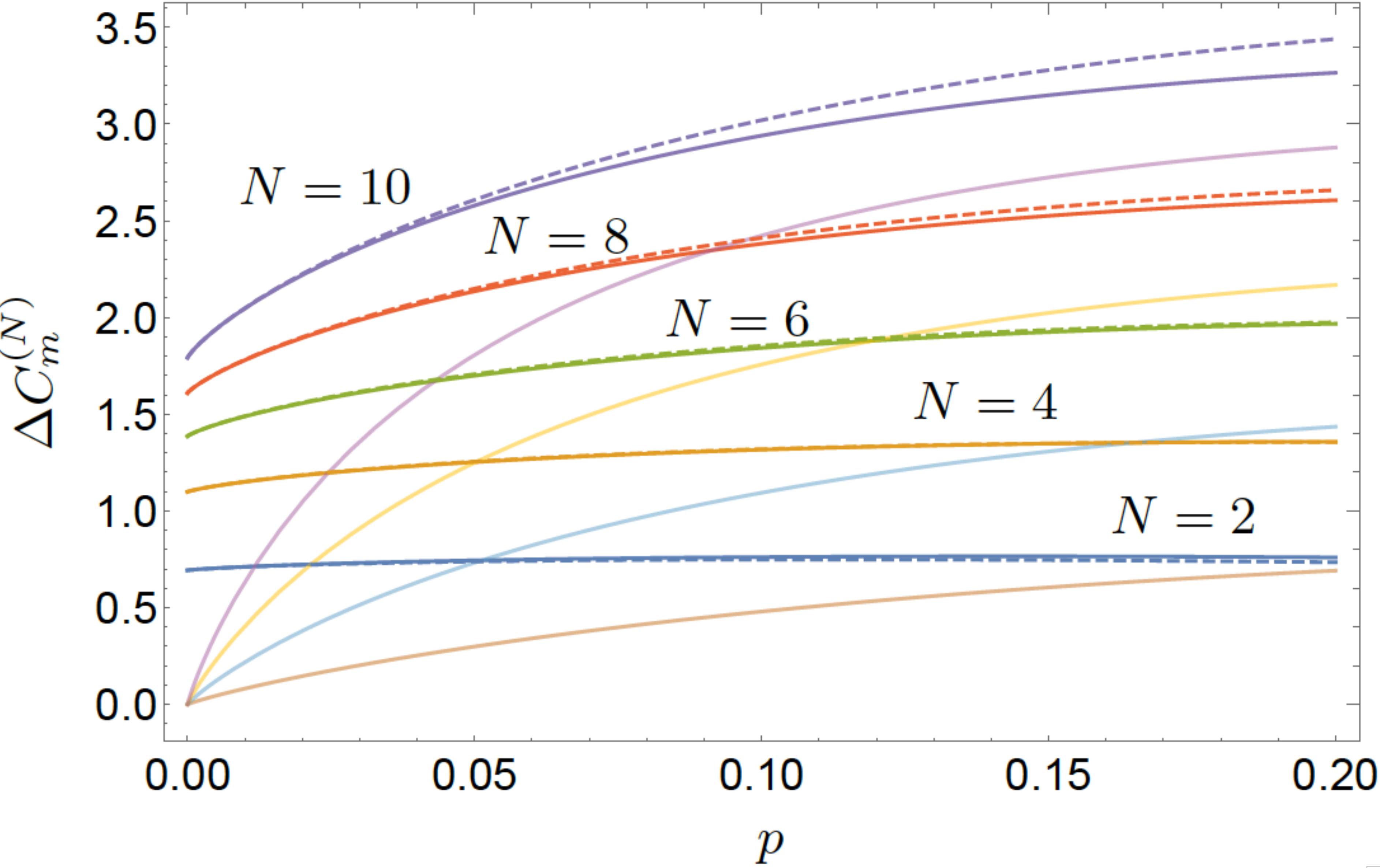}
\caption{Plot of the $p$-dependence of mutual coherence gain $\Delta C_{m}^{(N)}>0$, Eq.~\eqref{eq-C-mut}, of $N$ TLS in pure initial states. The plot shows curves in region of $p\ll 1$, again parametrized by $N$, as in Fig.~\ref{fig-delta-E-p} (the color code being the same). The corresponding approximations are shown as the dashed lines for even $N$ only. These results suggest, that the good match of the approximations holds in a comparable or larger interval of $p$, compared to the case of $\Delta C^{(N)}$ shown in Fig.~\ref{fig-delta-C-p}. Qualitative comparison with Fig.~\ref{fig-delta-C-p} reveals, that the mutual coherence $\Delta C_m^{(N)}$ is a {\it non-decreasing} function of $N$ and $p$ as well, on contrary to $\Delta C^{(N)}$.}
\label{fig-delta-Cm-p}
\end{figure}

This short appendix presents the behavior of $\Delta{E}^{(N)}$, $\Delta{C}^{(N)}$, $\Delta{C}_m^{(N)}$, and their corresponding approximations \eqref{eq-approx-DE}-\eqref{eq-pure-DCm-approx} (dashed lines in Figs.~\ref{fig-delta-E-p}-\ref{fig-delta-Cm-p},) from a complementary perspective. Namely, we present here their $p$-dependence of the above mentioned quantities with the number of TLS $N$ playing the role of the parameter. Figure~\ref{fig-delta-C-p} shows the main reason why we have restricted our attention to $p\ll 1$ region when dealing with $\Delta{C}^{(N)}>0$, as the approximation \eqref{eq-pure-DC-approx} is accurate in this region only. In turn, we restrict our attention to this region for the rest of the quantities, as well.  

On contrary, the approximations of the energy gain $\Delta{E}^{(N)}>0$, Fig.~\ref{fig-delta-E-p}, and mutual coherence gain $\Delta{C}_m^{(N)}>0$, Fig.~\ref{fig-delta-Cm-p}, reveal good match with the exact values in a wider region of $p$ values. 

\clearpage
\bibliography{apssamp}

\providecommand{\noopsort}[1]{}\providecommand{\singleletter}[1]{#1}%
\begin{thebibliography}{45}%
\makeatletter
\providecommand \@ifxundefined [1]{%
 \@ifx{#1\undefined}
}%
\providecommand \@ifnum [1]{%
 \ifnum #1\expandafter \@firstoftwo
 \else \expandafter \@secondoftwo
 \fi
}%
\providecommand \@ifx [1]{%
 \ifx #1\expandafter \@firstoftwo
 \else \expandafter \@secondoftwo
 \fi
}%
\providecommand \natexlab [1]{#1}%
\providecommand \enquote  [1]{``#1''}%
\providecommand \bibnamefont  [1]{#1}%
\providecommand \bibfnamefont [1]{#1}%
\providecommand \citenamefont [1]{#1}%
\providecommand \href@noop [0]{\@secondoftwo}%
\providecommand \href [0]{\begingroup \@sanitize@url \@href}%
\providecommand \@href[1]{\@@startlink{#1}\@@href}%
\providecommand \@@href[1]{\endgroup#1\@@endlink}%
\providecommand \@sanitize@url [0]{\catcode `\\12\catcode `\$12\catcode
  `\&12\catcode `\#12\catcode `\^12\catcode `\_12\catcode `\%12\relax}%
\providecommand \@@startlink[1]{}%
\providecommand \@@endlink[0]{}%
\providecommand \url  [0]{\begingroup\@sanitize@url \@url }%
\providecommand \@url [1]{\endgroup\@href {#1}{\urlprefix }}%
\providecommand \urlprefix  [0]{URL }%
\providecommand \Eprint [0]{\href }%
\providecommand \doibase [0]{https://doi.org/}%
\providecommand \selectlanguage [0]{\@gobble}%
\providecommand \bibinfo  [0]{\@secondoftwo}%
\providecommand \bibfield  [0]{\@secondoftwo}%
\providecommand \translation [1]{[#1]}%
\providecommand \BibitemOpen [0]{}%
\providecommand \bibitemStop [0]{}%
\providecommand \bibitemNoStop [0]{.\EOS\space}%
\providecommand \EOS [0]{\spacefactor3000\relax}%
\providecommand \BibitemShut  [1]{\csname bibitem#1\endcsname}%
\let\auto@bib@innerbib\@empty
\bibitem [{\citenamefont {Slussarenko}\ and\ \citenamefont
  {Pryde}(2019)}]{SlussarenkoAPR2019}%
  \BibitemOpen
  \bibfield  {author} {\bibinfo {author} {\bibfnamefont {S.}~\bibnamefont
  {Slussarenko}}\ and\ \bibinfo {author} {\bibfnamefont {G.~J.}\ \bibnamefont
  {Pryde}},\ }\bibfield  {title} {\bibinfo {title} {Photonic quantum
  information processing: A concise review},\ }\href
  {https://doi.org/10.1063/1.5115814} {\bibfield  {journal} {\bibinfo
  {journal} {Applied Physics Reviews}\ }\textbf {\bibinfo {volume} {6}},\
  \bibinfo {pages} {041303} (\bibinfo {year} {2019})}\BibitemShut {NoStop}%
\bibitem [{\citenamefont {Bruzewicz}\ \emph {et~al.}(2019)\citenamefont
  {Bruzewicz}, \citenamefont {Chiaverini}, \citenamefont {McConnell},\ and\
  \citenamefont {Sage}}]{BruzewiczAPR2019}%
  \BibitemOpen
  \bibfield  {author} {\bibinfo {author} {\bibfnamefont {C.~D.}\ \bibnamefont
  {Bruzewicz}}, \bibinfo {author} {\bibfnamefont {J.}~\bibnamefont
  {Chiaverini}}, \bibinfo {author} {\bibfnamefont {R.}~\bibnamefont
  {McConnell}},\ and\ \bibinfo {author} {\bibfnamefont {J.~M.}\ \bibnamefont
  {Sage}},\ }\bibfield  {title} {\bibinfo {title} {Trapped-ion quantum
  computing: Progress and challenges},\ }\href
  {https://doi.org/10.1063/1.5088164} {\bibfield  {journal} {\bibinfo
  {journal} {Applied Physics Reviews}\ }\textbf {\bibinfo {volume} {6}},\
  \bibinfo {pages} {021314} (\bibinfo {year} {2019})}\BibitemShut {NoStop}%
\bibitem [{\citenamefont {Wendin}(2017)}]{Wendin_2017}%
  \BibitemOpen
  \bibfield  {author} {\bibinfo {author} {\bibfnamefont {G.}~\bibnamefont
  {Wendin}},\ }\bibfield  {title} {\bibinfo {title} {Quantum information
  processing with superconducting circuits: a review},\ }\href
  {https://doi.org/10.1088/1361-6633/aa7e1a} {\bibfield  {journal} {\bibinfo
  {journal} {Reports on Progress in Physics}\ }\textbf {\bibinfo {volume}
  {80}},\ \bibinfo {pages} {106001} (\bibinfo {year} {2017})}\BibitemShut
  {NoStop}%
\bibitem [{\citenamefont {Degen}\ \emph {et~al.}(2017)\citenamefont {Degen},
  \citenamefont {Reinhard},\ and\ \citenamefont
  {Cappellaro}}]{DegenRevModPhys2017}%
  \BibitemOpen
  \bibfield  {author} {\bibinfo {author} {\bibfnamefont {C.~L.}\ \bibnamefont
  {Degen}}, \bibinfo {author} {\bibfnamefont {F.}~\bibnamefont {Reinhard}},\
  and\ \bibinfo {author} {\bibfnamefont {P.}~\bibnamefont {Cappellaro}},\
  }\bibfield  {title} {\bibinfo {title} {Quantum sensing},\ }\href
  {https://doi.org/10.1103/RevModPhys.89.035002} {\bibfield  {journal}
  {\bibinfo  {journal} {Rev. Mod. Phys.}\ }\textbf {\bibinfo {volume} {89}},\
  \bibinfo {pages} {035002} (\bibinfo {year} {2017})}\BibitemShut {NoStop}%
\bibitem [{\citenamefont {Anders}\ and\ \citenamefont
  {Esposito}(2017)}]{Anders_2017}%
  \BibitemOpen
  \bibfield  {author} {\bibinfo {author} {\bibfnamefont {J.}~\bibnamefont
  {Anders}}\ and\ \bibinfo {author} {\bibfnamefont {M.}~\bibnamefont
  {Esposito}},\ }\bibfield  {title} {\bibinfo {title} {Focus on quantum
  thermodynamics},\ }\href {https://doi.org/10.1088/1367-2630/19/1/010201}
  {\bibfield  {journal} {\bibinfo  {journal} {New Journal of Physics}\ }\textbf
  {\bibinfo {volume} {19}},\ \bibinfo {pages} {010201} (\bibinfo {year}
  {2017})}\BibitemShut {NoStop}%
\bibitem [{\citenamefont {Goold}\ \emph {et~al.}(2016)\citenamefont {Goold},
  \citenamefont {Huber}, \citenamefont {Riera}, \citenamefont {del Rio},\ and\
  \citenamefont {Skrzypczyk}}]{Goold_2016}%
  \BibitemOpen
  \bibfield  {author} {\bibinfo {author} {\bibfnamefont {J.}~\bibnamefont
  {Goold}}, \bibinfo {author} {\bibfnamefont {M.}~\bibnamefont {Huber}},
  \bibinfo {author} {\bibfnamefont {A.}~\bibnamefont {Riera}}, \bibinfo
  {author} {\bibfnamefont {L.}~\bibnamefont {del Rio}},\ and\ \bibinfo {author}
  {\bibfnamefont {P.}~\bibnamefont {Skrzypczyk}},\ }\bibfield  {title}
  {\bibinfo {title} {The role of quantum information in
  thermodynamics{\textemdash}a topical review},\ }\href
  {https://doi.org/10.1088/1751-8113/49/14/143001} {\bibfield  {journal}
  {\bibinfo  {journal} {Journal of Physics A: Mathematical and Theoretical}\
  }\textbf {\bibinfo {volume} {49}},\ \bibinfo {pages} {143001} (\bibinfo
  {year} {2016})}\BibitemShut {NoStop}%
\bibitem [{\citenamefont {Streltsov}\ \emph {et~al.}(2017)\citenamefont
  {Streltsov}, \citenamefont {Adesso},\ and\ \citenamefont
  {Plenio}}]{StreltsovRevModPhys2017}%
  \BibitemOpen
  \bibfield  {author} {\bibinfo {author} {\bibfnamefont {A.}~\bibnamefont
  {Streltsov}}, \bibinfo {author} {\bibfnamefont {G.}~\bibnamefont {Adesso}},\
  and\ \bibinfo {author} {\bibfnamefont {M.~B.}\ \bibnamefont {Plenio}},\
  }\bibfield  {title} {\bibinfo {title} {Colloquium: Quantum coherence as a
  resource},\ }\href {https://doi.org/10.1103/RevModPhys.89.041003} {\bibfield
  {journal} {\bibinfo  {journal} {Rev. Mod. Phys.}\ }\textbf {\bibinfo {volume}
  {89}},\ \bibinfo {pages} {041003} (\bibinfo {year} {2017})}\BibitemShut
  {NoStop}%
\bibitem [{\citenamefont {Liu}\ and\ \citenamefont {Zhou}(2019)}]{LiuPRL2019}%
  \BibitemOpen
  \bibfield  {author} {\bibinfo {author} {\bibfnamefont {C.~L.}\ \bibnamefont
  {Liu}}\ and\ \bibinfo {author} {\bibfnamefont {D.~L.}\ \bibnamefont {Zhou}},\
  }\bibfield  {title} {\bibinfo {title} {Deterministic coherence
  distillation},\ }\href {https://doi.org/10.1103/PhysRevLett.123.070402}
  {\bibfield  {journal} {\bibinfo  {journal} {Phys. Rev. Lett.}\ }\textbf
  {\bibinfo {volume} {123}},\ \bibinfo {pages} {070402} (\bibinfo {year}
  {2019})}\BibitemShut {NoStop}%
\bibitem [{\citenamefont {Wu}\ \emph {et~al.}(2020)\citenamefont {Wu},
  \citenamefont {Theurer}, \citenamefont {Xiang}, \citenamefont {Li},
  \citenamefont {Guo}, \citenamefont {Plenio},\ and\ \citenamefont
  {Streltsov}}]{WuNPJ2020}%
  \BibitemOpen
  \bibfield  {author} {\bibinfo {author} {\bibfnamefont {K.-D.}\ \bibnamefont
  {Wu}}, \bibinfo {author} {\bibfnamefont {T.}~\bibnamefont {Theurer}},
  \bibinfo {author} {\bibfnamefont {G.-Y.}\ \bibnamefont {Xiang}}, \bibinfo
  {author} {\bibfnamefont {C.-F.}\ \bibnamefont {Li}}, \bibinfo {author}
  {\bibfnamefont {G.-C.}\ \bibnamefont {Guo}}, \bibinfo {author} {\bibfnamefont
  {M.~B.}\ \bibnamefont {Plenio}},\ and\ \bibinfo {author} {\bibfnamefont
  {A.}~\bibnamefont {Streltsov}},\ }\bibfield  {title} {\bibinfo {title}
  {Quantum coherence and state conversion: theory and experiment},\ }\href
  {https://doi.org/10.1038/s41534-020-0250-z} {\bibfield  {journal} {\bibinfo
  {journal} {npj Quantum Information}\ }\textbf {\bibinfo {volume} {6}},\
  \bibinfo {pages} {22} (\bibinfo {year} {2020})}\BibitemShut {NoStop}%
\bibitem [{\citenamefont {Pang}\ and\ \citenamefont
  {Zhao}(2020)}]{PangQIP2020}%
  \BibitemOpen
  \bibfield  {author} {\bibinfo {author} {\bibfnamefont {Z.-Y.}\ \bibnamefont
  {Pang}}\ and\ \bibinfo {author} {\bibfnamefont {M.-J.}\ \bibnamefont
  {Zhao}},\ }\bibfield  {title} {\bibinfo {title} {Probabilistic coherence
  distillation with assisted setting},\ }\href
  {https://doi.org/10.1007/s11128-020-02857-5} {\bibfield  {journal} {\bibinfo
  {journal} {Quantum Information Processing}\ }\textbf {\bibinfo {volume}
  {19}},\ \bibinfo {pages} {363} (\bibinfo {year} {2020})}\BibitemShut
  {NoStop}%
\bibitem [{\citenamefont {St{\'{a}}rek}\ \emph {et~al.}(2021)\citenamefont
  {St{\'{a}}rek}, \citenamefont {Mi{\v{c}}uda}, \citenamefont
  {Kol{\'{a}}{\v{r}}}, \citenamefont {Filip},\ and\ \citenamefont
  {Fiur{\'{a}}{\v{s}}ek}}]{Starek2021}%
  \BibitemOpen
  \bibfield  {author} {\bibinfo {author} {\bibfnamefont {R.}~\bibnamefont
  {St{\'{a}}rek}}, \bibinfo {author} {\bibfnamefont {M.}~\bibnamefont
  {Mi{\v{c}}uda}}, \bibinfo {author} {\bibfnamefont {M.}~\bibnamefont
  {Kol{\'{a}}{\v{r}}}}, \bibinfo {author} {\bibfnamefont {R.}~\bibnamefont
  {Filip}},\ and\ \bibinfo {author} {\bibfnamefont {J.}~\bibnamefont
  {Fiur{\'{a}}{\v{s}}ek}},\ }\bibfield  {title} {\bibinfo {title} {Experimental
  demonstration of optimal probabilistic enhancement of quantum coherence},\
  }\href {https://doi.org/10.1088/2058-9565/ac10ef} {\bibfield  {journal}
  {\bibinfo  {journal} {Quantum Science and Technology}\ }\textbf {\bibinfo
  {volume} {6}},\ \bibinfo {pages} {045010} (\bibinfo {year}
  {2021})}\BibitemShut {NoStop}%
\bibitem [{\citenamefont {Guarnieri}\ \emph {et~al.}(2018)\citenamefont
  {Guarnieri}, \citenamefont {Kol\'a\ifmmode~\check{r}\else \v{r}\fi{}},\ and\
  \citenamefont {Filip}}]{giacomoPRL2018}%
  \BibitemOpen
  \bibfield  {author} {\bibinfo {author} {\bibfnamefont {G.}~\bibnamefont
  {Guarnieri}}, \bibinfo {author} {\bibfnamefont {M.}~\bibnamefont
  {Kol\'a\ifmmode~\check{r}\else \v{r}\fi{}}},\ and\ \bibinfo {author}
  {\bibfnamefont {R.}~\bibnamefont {Filip}},\ }\bibfield  {title} {\bibinfo
  {title} {Steady-state coherences by composite system-bath interactions},\
  }\href {https://doi.org/https://doi.org/10.1103/PhysRevLett.121.070401}
  {\bibfield  {journal} {\bibinfo  {journal} {Phys. Rev. Lett.}\ }\textbf
  {\bibinfo {volume} {121}},\ \bibinfo {pages} {070401} (\bibinfo {year}
  {2018})}\BibitemShut {NoStop}%
\bibitem [{\citenamefont {Román-Ancheyta}\ \emph {et~al.}(2020)\citenamefont
  {Román-Ancheyta}, \citenamefont {Kolář}, \citenamefont {Guarnieri},\ and\
  \citenamefont {Filip}}]{romanancheyta2020enhanced}%
  \BibitemOpen
  \bibfield  {author} {\bibinfo {author} {\bibfnamefont {R.}~\bibnamefont
  {Román-Ancheyta}}, \bibinfo {author} {\bibfnamefont {M.}~\bibnamefont
  {Kolář}}, \bibinfo {author} {\bibfnamefont {G.}~\bibnamefont {Guarnieri}},\
  and\ \bibinfo {author} {\bibfnamefont {R.}~\bibnamefont {Filip}},\
  }\href@noop {} {\bibinfo {title} {Enhanced steady-state coherences via
  repeated system-bath interactions}} (\bibinfo {year} {2020}),\ \Eprint
  {https://arxiv.org/abs/2008.05200} {arXiv:2008.05200 [quant-ph]} \BibitemShut
  {NoStop}%
\bibitem [{\citenamefont {Purkayastha}\ \emph {et~al.}(2020)\citenamefont
  {Purkayastha}, \citenamefont {Guarnieri}, \citenamefont {Mitchison},
  \citenamefont {Filip},\ and\ \citenamefont {Goold}}]{ArchakNPJ2020}%
  \BibitemOpen
  \bibfield  {author} {\bibinfo {author} {\bibfnamefont {A.}~\bibnamefont
  {Purkayastha}}, \bibinfo {author} {\bibfnamefont {G.}~\bibnamefont
  {Guarnieri}}, \bibinfo {author} {\bibfnamefont {M.~T.}\ \bibnamefont
  {Mitchison}}, \bibinfo {author} {\bibfnamefont {R.}~\bibnamefont {Filip}},\
  and\ \bibinfo {author} {\bibfnamefont {J.}~\bibnamefont {Goold}},\ }\bibfield
   {title} {\bibinfo {title} {Tunable phonon-induced steady-state coherence in
  a double-quantum-dot charge qubit},\ }\href
  {https://doi.org/10.1038/s41534-020-0256-6} {\bibfield  {journal} {\bibinfo
  {journal} {npj Quantum Information}\ }\textbf {\bibinfo {volume} {6}},\
  \bibinfo {pages} {22} (\bibinfo {year} {2020})}\BibitemShut {NoStop}%
\bibitem [{\citenamefont {Guarnieri}\ \emph {et~al.}(2020)\citenamefont
  {Guarnieri}, \citenamefont {Morrone}, \citenamefont {Çakmak}, \citenamefont
  {Plastina},\ and\ \citenamefont {Campbell}}]{GUARNIERIPLA2020}%
  \BibitemOpen
  \bibfield  {author} {\bibinfo {author} {\bibfnamefont {G.}~\bibnamefont
  {Guarnieri}}, \bibinfo {author} {\bibfnamefont {D.}~\bibnamefont {Morrone}},
  \bibinfo {author} {\bibfnamefont {B.}~\bibnamefont {Çakmak}}, \bibinfo
  {author} {\bibfnamefont {F.}~\bibnamefont {Plastina}},\ and\ \bibinfo
  {author} {\bibfnamefont {S.}~\bibnamefont {Campbell}},\ }\bibfield  {title}
  {\bibinfo {title} {Non-equilibrium steady-states of memoryless quantum
  collision models},\ }\href
  {https://doi.org/https://doi.org/10.1016/j.physleta.2020.126576} {\bibfield
  {journal} {\bibinfo  {journal} {Physics Letters A}\ }\textbf {\bibinfo
  {volume} {384}},\ \bibinfo {pages} {126576} (\bibinfo {year}
  {2020})}\BibitemShut {NoStop}%
\bibitem [{\citenamefont {Gherardini}\ \emph {et~al.}(2020)\citenamefont
  {Gherardini}, \citenamefont {Campaioli}, \citenamefont {Caruso},\ and\
  \citenamefont {Binder}}]{Campaioli2020}%
  \BibitemOpen
  \bibfield  {author} {\bibinfo {author} {\bibfnamefont {S.}~\bibnamefont
  {Gherardini}}, \bibinfo {author} {\bibfnamefont {F.}~\bibnamefont
  {Campaioli}}, \bibinfo {author} {\bibfnamefont {F.}~\bibnamefont {Caruso}},\
  and\ \bibinfo {author} {\bibfnamefont {F.~C.}\ \bibnamefont {Binder}},\
  }\bibfield  {title} {\bibinfo {title} {Stabilizing open quantum batteries by
  sequential measurements},\ }\href
  {https://doi.org/10.1103/PhysRevResearch.2.013095} {\bibfield  {journal}
  {\bibinfo  {journal} {Phys. Rev. Research}\ }\textbf {\bibinfo {volume}
  {2}},\ \bibinfo {pages} {013095} (\bibinfo {year} {2020})}\BibitemShut
  {NoStop}%
\bibitem [{\citenamefont {Manatuly}\ \emph {et~al.}(2019)\citenamefont
  {Manatuly}, \citenamefont {Niedenzu}, \citenamefont {Rom\'an-Ancheyta},
  \citenamefont {\ifmmode~\mbox{\c{C}}\else \c{C}\fi{}akmak}, \citenamefont
  {M\"ustecapl\ifmmode \imath \else \i \fi{}o\ifmmode~\breve{g}\else
  \u{g}\fi{}lu},\ and\ \citenamefont {Kurizki}}]{ancheytaPRE2019}%
  \BibitemOpen
  \bibfield  {author} {\bibinfo {author} {\bibfnamefont {A.}~\bibnamefont
  {Manatuly}}, \bibinfo {author} {\bibfnamefont {W.}~\bibnamefont {Niedenzu}},
  \bibinfo {author} {\bibfnamefont {R.}~\bibnamefont {Rom\'an-Ancheyta}},
  \bibinfo {author} {\bibfnamefont {B.}~\bibnamefont
  {\ifmmode~\mbox{\c{C}}\else \c{C}\fi{}akmak}}, \bibinfo {author}
  {\bibfnamefont {O.~E.}\ \bibnamefont {M\"ustecapl\ifmmode \imath \else \i
  \fi{}o\ifmmode~\breve{g}\else \u{g}\fi{}lu}},\ and\ \bibinfo {author}
  {\bibfnamefont {G.}~\bibnamefont {Kurizki}},\ }\bibfield  {title} {\bibinfo
  {title} {Collectively enhanced thermalization via multiqubit collisions},\
  }\href {https://doi.org/10.1103/PhysRevE.99.042145} {\bibfield  {journal}
  {\bibinfo  {journal} {Phys. Rev. E}\ }\textbf {\bibinfo {volume} {99}},\
  \bibinfo {pages} {042145} (\bibinfo {year} {2019})}\BibitemShut {NoStop}%
\bibitem [{\citenamefont {Klatzow}\ \emph {et~al.}(2019)\citenamefont
  {Klatzow}, \citenamefont {Becker}, \citenamefont {Ledingham}, \citenamefont
  {Weinzetl}, \citenamefont {Kaczmarek}, \citenamefont {Saunders},
  \citenamefont {Nunn}, \citenamefont {Walmsley}, \citenamefont {Uzdin},\ and\
  \citenamefont {Poem}}]{KlatzowPRL2019}%
  \BibitemOpen
  \bibfield  {author} {\bibinfo {author} {\bibfnamefont {J.}~\bibnamefont
  {Klatzow}}, \bibinfo {author} {\bibfnamefont {J.~N.}\ \bibnamefont {Becker}},
  \bibinfo {author} {\bibfnamefont {P.~M.}\ \bibnamefont {Ledingham}}, \bibinfo
  {author} {\bibfnamefont {C.}~\bibnamefont {Weinzetl}}, \bibinfo {author}
  {\bibfnamefont {K.~T.}\ \bibnamefont {Kaczmarek}}, \bibinfo {author}
  {\bibfnamefont {D.~J.}\ \bibnamefont {Saunders}}, \bibinfo {author}
  {\bibfnamefont {J.}~\bibnamefont {Nunn}}, \bibinfo {author} {\bibfnamefont
  {I.~A.}\ \bibnamefont {Walmsley}}, \bibinfo {author} {\bibfnamefont
  {R.}~\bibnamefont {Uzdin}},\ and\ \bibinfo {author} {\bibfnamefont
  {E.}~\bibnamefont {Poem}},\ }\bibfield  {title} {\bibinfo {title}
  {Experimental demonstration of quantum effects in the operation of
  microscopic heat engines},\ }\href
  {https://doi.org/10.1103/PhysRevLett.122.110601} {\bibfield  {journal}
  {\bibinfo  {journal} {Phys. Rev. Lett.}\ }\textbf {\bibinfo {volume} {122}},\
  \bibinfo {pages} {110601} (\bibinfo {year} {2019})}\BibitemShut {NoStop}%
\bibitem [{\citenamefont {Kol\'a\ifmmode~\check{r}\else \v{r}\fi{}}\ \emph
  {et~al.}(2017)\citenamefont {Kol\'a\ifmmode~\check{r}\else \v{r}\fi{}},
  \citenamefont {Ryabov},\ and\ \citenamefont {Filip}}]{kolarPRA2017}%
  \BibitemOpen
  \bibfield  {author} {\bibinfo {author} {\bibfnamefont {M.}~\bibnamefont
  {Kol\'a\ifmmode~\check{r}\else \v{r}\fi{}}}, \bibinfo {author} {\bibfnamefont
  {A.}~\bibnamefont {Ryabov}},\ and\ \bibinfo {author} {\bibfnamefont
  {R.}~\bibnamefont {Filip}},\ }\bibfield  {title} {\bibinfo {title}
  {Optomechanical oscillator controlled by variation in its heat bath
  temperature},\ }\href {https://doi.org/10.1103/PhysRevA.95.042105} {\bibfield
   {journal} {\bibinfo  {journal} {Phys. Rev. A}\ }\textbf {\bibinfo {volume}
  {95}},\ \bibinfo {pages} {042105} (\bibinfo {year} {2017})}\BibitemShut
  {NoStop}%
\bibitem [{\citenamefont {Alicki}\ and\ \citenamefont
  {Fannes}(2013)}]{alicki2013}%
  \BibitemOpen
  \bibfield  {author} {\bibinfo {author} {\bibfnamefont {R.}~\bibnamefont
  {Alicki}}\ and\ \bibinfo {author} {\bibfnamefont {M.}~\bibnamefont
  {Fannes}},\ }\bibfield  {title} {\bibinfo {title} {Entanglement boost for
  extractable work from ensembles of quantum batteries},\ }\href
  {https://doi.org/https://doi.org/10.1103/PhysRevE.87.042123} {\bibfield
  {journal} {\bibinfo  {journal} {Phys. Rev. E}\ }\textbf {\bibinfo {volume}
  {87}},\ \bibinfo {pages} {042123} (\bibinfo {year} {2013})}\BibitemShut
  {NoStop}%
\bibitem [{\citenamefont {Binder}\ \emph {et~al.}(2015)\citenamefont {Binder},
  \citenamefont {Vinjanampathy}, \citenamefont {Modi},\ and\ \citenamefont
  {Goold}}]{binderNJP2015}%
  \BibitemOpen
  \bibfield  {author} {\bibinfo {author} {\bibfnamefont {F.}~\bibnamefont
  {Binder}}, \bibinfo {author} {\bibfnamefont {S.}~\bibnamefont
  {Vinjanampathy}}, \bibinfo {author} {\bibfnamefont {K.}~\bibnamefont
  {Modi}},\ and\ \bibinfo {author} {\bibfnamefont {J.}~\bibnamefont {Goold}},\
  }\bibfield  {title} {{\selectlanguage {English}\bibinfo {title} {Quantacell:
  powerful charging of quantum batteries}},\ }\href
  {https://doi.org/10.1088/1367-2630/17/7/075015} {\bibfield  {journal}
  {\bibinfo  {journal} {New Journal of Physics}\ }\textbf {\bibinfo {volume}
  {17}} (\bibinfo {year} {2015})}\BibitemShut {NoStop}%
\bibitem [{\citenamefont {Pirmoradian}\ and\ \citenamefont
  {M\o{}lmer}(2019)}]{Pirmoradian2019}%
  \BibitemOpen
  \bibfield  {author} {\bibinfo {author} {\bibfnamefont {F.}~\bibnamefont
  {Pirmoradian}}\ and\ \bibinfo {author} {\bibfnamefont {K.}~\bibnamefont
  {M\o{}lmer}},\ }\bibfield  {title} {\bibinfo {title} {Aging of a quantum
  battery},\ }\href {https://doi.org/10.1103/PhysRevA.100.043833} {\bibfield
  {journal} {\bibinfo  {journal} {Phys. Rev. A}\ }\textbf {\bibinfo {volume}
  {100}},\ \bibinfo {pages} {043833} (\bibinfo {year} {2019})}\BibitemShut
  {NoStop}%
\bibitem [{\citenamefont {Santos}\ \emph {et~al.}(2019)\citenamefont {Santos},
  \citenamefont {\ifmmode~\mbox{\c{C}}\else \c{C}\fi{}akmak}, \citenamefont
  {Campbell},\ and\ \citenamefont {Zinner}}]{santos2019}%
  \BibitemOpen
  \bibfield  {author} {\bibinfo {author} {\bibfnamefont {A.~C.}\ \bibnamefont
  {Santos}}, \bibinfo {author} {\bibfnamefont {B.}~\bibnamefont
  {\ifmmode~\mbox{\c{C}}\else \c{C}\fi{}akmak}}, \bibinfo {author}
  {\bibfnamefont {S.}~\bibnamefont {Campbell}},\ and\ \bibinfo {author}
  {\bibfnamefont {N.~T.}\ \bibnamefont {Zinner}},\ }\bibfield  {title}
  {\bibinfo {title} {Stable adiabatic quantum batteries},\ }\href
  {https://doi.org/10.1103/PhysRevE.100.032107} {\bibfield  {journal} {\bibinfo
   {journal} {Phys. Rev. E}\ }\textbf {\bibinfo {volume} {100}},\ \bibinfo
  {pages} {032107} (\bibinfo {year} {2019})}\BibitemShut {NoStop}%
\bibitem [{\citenamefont {Campaioli}\ \emph {et~al.}(2017)\citenamefont
  {Campaioli}, \citenamefont {Pollock}, \citenamefont {Binder}, \citenamefont
  {C\'eleri}, \citenamefont {Goold}, \citenamefont {Vinjanampathy},\ and\
  \citenamefont {Modi}}]{campaioliPRL2017}%
  \BibitemOpen
  \bibfield  {author} {\bibinfo {author} {\bibfnamefont {F.}~\bibnamefont
  {Campaioli}}, \bibinfo {author} {\bibfnamefont {F.~A.}\ \bibnamefont
  {Pollock}}, \bibinfo {author} {\bibfnamefont {F.~C.}\ \bibnamefont {Binder}},
  \bibinfo {author} {\bibfnamefont {L.}~\bibnamefont {C\'eleri}}, \bibinfo
  {author} {\bibfnamefont {J.}~\bibnamefont {Goold}}, \bibinfo {author}
  {\bibfnamefont {S.}~\bibnamefont {Vinjanampathy}},\ and\ \bibinfo {author}
  {\bibfnamefont {K.}~\bibnamefont {Modi}},\ }\bibfield  {title} {\bibinfo
  {title} {Enhancing the charging power of quantum batteries},\ }\href
  {https://doi.org/https://doi.org/10.1103/PhysRevLett.118.150601} {\bibfield
  {journal} {\bibinfo  {journal} {Phys. Rev. Lett.}\ }\textbf {\bibinfo
  {volume} {118}},\ \bibinfo {pages} {150601} (\bibinfo {year}
  {2017})}\BibitemShut {NoStop}%
\bibitem [{\citenamefont {Gumberidze}\ \emph {et~al.}(2019)\citenamefont
  {Gumberidze}, \citenamefont {Kolář},\ and\ \citenamefont
  {Filip}}]{Gumberidze2019}%
  \BibitemOpen
  \bibfield  {author} {\bibinfo {author} {\bibfnamefont {M.}~\bibnamefont
  {Gumberidze}}, \bibinfo {author} {\bibfnamefont {M.}~\bibnamefont
  {Kolář}},\ and\ \bibinfo {author} {\bibfnamefont {R.}~\bibnamefont
  {Filip}},\ }\bibfield  {title} {\bibinfo {title} {Measurement induced
  synthesis of coherent quantum batteries},\ }\href
  {https://doi.org/10.1038/s41598-019-56158-8} {\bibfield  {journal} {\bibinfo
  {journal} {Scientific Reports}\ }\textbf {\bibinfo {volume} {9}},\ \bibinfo
  {pages} {19628} (\bibinfo {year} {2019})}\BibitemShut {NoStop}%
\bibitem [{\citenamefont {Xi}\ \emph {et~al.}(2015)\citenamefont {Xi},
  \citenamefont {Li},\ and\ \citenamefont {Fan}}]{XiSciRep2015}%
  \BibitemOpen
  \bibfield  {author} {\bibinfo {author} {\bibfnamefont {Z.}~\bibnamefont
  {Xi}}, \bibinfo {author} {\bibfnamefont {Y.}~\bibnamefont {Li}},\ and\
  \bibinfo {author} {\bibfnamefont {H.}~\bibnamefont {Fan}},\ }\bibfield
  {title} {\bibinfo {title} {Quantum coherence and correlations in quantum
  system},\ }\href {https://doi.org/10.1038/srep10922} {\bibfield  {journal}
  {\bibinfo  {journal} {Scientific Reports}\ }\textbf {\bibinfo {volume} {5}},\
  \bibinfo {pages} {10922} (\bibinfo {year} {2015})}\BibitemShut {NoStop}%
\bibitem [{\citenamefont {Guo}\ and\ \citenamefont
  {Goswami}(2017)}]{GuoPhysRevA2017}%
  \BibitemOpen
  \bibfield  {author} {\bibinfo {author} {\bibfnamefont {Y.}~\bibnamefont
  {Guo}}\ and\ \bibinfo {author} {\bibfnamefont {S.}~\bibnamefont {Goswami}},\
  }\bibfield  {title} {\bibinfo {title} {Discordlike correlation of bipartite
  coherence},\ }\href {https://doi.org/10.1103/PhysRevA.95.062340} {\bibfield
  {journal} {\bibinfo  {journal} {Phys. Rev. A}\ }\textbf {\bibinfo {volume}
  {95}},\ \bibinfo {pages} {062340} (\bibinfo {year} {2017})}\BibitemShut
  {NoStop}%
\bibitem [{\citenamefont {Wang}\ \emph {et~al.}(2017)\citenamefont {Wang},
  \citenamefont {Yue}, \citenamefont {Yu}, \citenamefont {Gao},\ and\
  \citenamefont {Qin}}]{WangSciRep2017}%
  \BibitemOpen
  \bibfield  {author} {\bibinfo {author} {\bibfnamefont {X.-L.}\ \bibnamefont
  {Wang}}, \bibinfo {author} {\bibfnamefont {Q.-L.}\ \bibnamefont {Yue}},
  \bibinfo {author} {\bibfnamefont {C.-H.}\ \bibnamefont {Yu}}, \bibinfo
  {author} {\bibfnamefont {F.}~\bibnamefont {Gao}},\ and\ \bibinfo {author}
  {\bibfnamefont {S.-J.}\ \bibnamefont {Qin}},\ }\bibfield  {title} {\bibinfo
  {title} {Relating quantum coherence and correlations with entropy-based
  measures},\ }\href {https://doi.org/10.1038/s41598-017-09332-9} {\bibfield
  {journal} {\bibinfo  {journal} {Scientific Reports}\ }\textbf {\bibinfo
  {volume} {7}},\ \bibinfo {pages} {12122} (\bibinfo {year}
  {2017})}\BibitemShut {NoStop}%
\bibitem [{\citenamefont {Tan}\ and\ \citenamefont
  {Jeong}(2018)}]{TanPhysRevLett2018}%
  \BibitemOpen
  \bibfield  {author} {\bibinfo {author} {\bibfnamefont {K.~C.}\ \bibnamefont
  {Tan}}\ and\ \bibinfo {author} {\bibfnamefont {H.}~\bibnamefont {Jeong}},\
  }\bibfield  {title} {\bibinfo {title} {Entanglement as the symmetric portion
  of correlated coherence},\ }\href
  {https://doi.org/10.1103/PhysRevLett.121.220401} {\bibfield  {journal}
  {\bibinfo  {journal} {Phys. Rev. Lett.}\ }\textbf {\bibinfo {volume} {121}},\
  \bibinfo {pages} {220401} (\bibinfo {year} {2018})}\BibitemShut {NoStop}%
\bibitem [{\citenamefont {Kraft}\ and\ \citenamefont
  {Piani}(2018)}]{Kraft_2018}%
  \BibitemOpen
  \bibfield  {author} {\bibinfo {author} {\bibfnamefont {T.}~\bibnamefont
  {Kraft}}\ and\ \bibinfo {author} {\bibfnamefont {M.}~\bibnamefont {Piani}},\
  }\bibfield  {title} {\bibinfo {title} {Genuine correlated coherence},\ }\href
  {https://doi.org/10.1088/1751-8121/aab8ad} {\bibfield  {journal} {\bibinfo
  {journal} {Journal of Physics A: Mathematical and Theoretical}\ }\textbf
  {\bibinfo {volume} {51}},\ \bibinfo {pages} {414013} (\bibinfo {year}
  {2018})}\BibitemShut {NoStop}%
\bibitem [{\citenamefont {Kammerlander}\ and\ \citenamefont
  {Anders}(2016)}]{kammerlander}%
  \BibitemOpen
  \bibfield  {author} {\bibinfo {author} {\bibfnamefont {P.}~\bibnamefont
  {Kammerlander}}\ and\ \bibinfo {author} {\bibfnamefont {J.}~\bibnamefont
  {Anders}},\ }\bibfield  {title} {\bibinfo {title} {Coherence and measurement
  in quantum thermodynamics},\ }\href
  {https://doi.org/https://doi.org/10.1038/srep22174} {\bibfield  {journal}
  {\bibinfo  {journal} {Scientific Reports}\ }\textbf {\bibinfo {volume} {6}},\
  \bibinfo {pages} {22174} (\bibinfo {year} {2016})}\BibitemShut {NoStop}%
\bibitem [{\citenamefont {Baumgratz}\ \emph {et~al.}(2014)\citenamefont
  {Baumgratz}, \citenamefont {Cramer},\ and\ \citenamefont
  {Plenio}}]{baumgratz}%
  \BibitemOpen
  \bibfield  {author} {\bibinfo {author} {\bibfnamefont {T.}~\bibnamefont
  {Baumgratz}}, \bibinfo {author} {\bibfnamefont {M.}~\bibnamefont {Cramer}},\
  and\ \bibinfo {author} {\bibfnamefont {M.~B.}\ \bibnamefont {Plenio}},\
  }\bibfield  {title} {\bibinfo {title} {Quantifying coherence},\ }\href
  {https://doi.org/https://doi.org/10.1103/PhysRevLett.113.140401} {\bibfield
  {journal} {\bibinfo  {journal} {Phys. Rev. Lett.}\ }\textbf {\bibinfo
  {volume} {113}},\ \bibinfo {pages} {140401} (\bibinfo {year}
  {2014})}\BibitemShut {NoStop}%
\bibitem [{\citenamefont {Yao}\ \emph {et~al.}(2015)\citenamefont {Yao},
  \citenamefont {Xiao}, \citenamefont {Ge},\ and\ \citenamefont
  {Sun}}]{YaoPRA2015}%
  \BibitemOpen
  \bibfield  {author} {\bibinfo {author} {\bibfnamefont {Y.}~\bibnamefont
  {Yao}}, \bibinfo {author} {\bibfnamefont {X.}~\bibnamefont {Xiao}}, \bibinfo
  {author} {\bibfnamefont {L.}~\bibnamefont {Ge}},\ and\ \bibinfo {author}
  {\bibfnamefont {C.~P.}\ \bibnamefont {Sun}},\ }\bibfield  {title} {\bibinfo
  {title} {Quantum coherence in multipartite systems},\ }\href
  {https://doi.org/10.1103/PhysRevA.92.022112} {\bibfield  {journal} {\bibinfo
  {journal} {Phys. Rev. A}\ }\textbf {\bibinfo {volume} {92}},\ \bibinfo
  {pages} {022112} (\bibinfo {year} {2015})}\BibitemShut {NoStop}%
\bibitem [{\citenamefont {Ma}\ \emph {et~al.}(2016)\citenamefont {Ma},
  \citenamefont {Yadin}, \citenamefont {Girolami}, \citenamefont {Vedral},\
  and\ \citenamefont {Gu}}]{MaPRL2016}%
  \BibitemOpen
  \bibfield  {author} {\bibinfo {author} {\bibfnamefont {J.}~\bibnamefont
  {Ma}}, \bibinfo {author} {\bibfnamefont {B.}~\bibnamefont {Yadin}}, \bibinfo
  {author} {\bibfnamefont {D.}~\bibnamefont {Girolami}}, \bibinfo {author}
  {\bibfnamefont {V.}~\bibnamefont {Vedral}},\ and\ \bibinfo {author}
  {\bibfnamefont {M.}~\bibnamefont {Gu}},\ }\bibfield  {title} {\bibinfo
  {title} {Converting coherence to quantum correlations},\ }\href
  {https://doi.org/10.1103/PhysRevLett.116.160407} {\bibfield  {journal}
  {\bibinfo  {journal} {Phys. Rev. Lett.}\ }\textbf {\bibinfo {volume} {116}},\
  \bibinfo {pages} {160407} (\bibinfo {year} {2016})}\BibitemShut {NoStop}%
\bibitem [{\citenamefont {Fang}\ \emph {et~al.}(2018)\citenamefont {Fang},
  \citenamefont {Wang}, \citenamefont {Lami}, \citenamefont {Regula},\ and\
  \citenamefont {Adesso}}]{adessoPRL2018}%
  \BibitemOpen
  \bibfield  {author} {\bibinfo {author} {\bibfnamefont {K.}~\bibnamefont
  {Fang}}, \bibinfo {author} {\bibfnamefont {X.}~\bibnamefont {Wang}}, \bibinfo
  {author} {\bibfnamefont {L.}~\bibnamefont {Lami}}, \bibinfo {author}
  {\bibfnamefont {B.}~\bibnamefont {Regula}},\ and\ \bibinfo {author}
  {\bibfnamefont {G.}~\bibnamefont {Adesso}},\ }\bibfield  {title} {\bibinfo
  {title} {Probabilistic distillation of quantum coherence},\ }\href
  {https://doi.org/10.1103/PhysRevLett.121.070404} {\bibfield  {journal}
  {\bibinfo  {journal} {Phys. Rev. Lett.}\ }\textbf {\bibinfo {volume} {121}},\
  \bibinfo {pages} {070404} (\bibinfo {year} {2018})}\BibitemShut {NoStop}%
\bibitem [{\citenamefont {Nielsen}\ and\ \citenamefont
  {Chuang}(2000)}]{NielsenBook}%
  \BibitemOpen
  \bibfield  {author} {\bibinfo {author} {\bibfnamefont {M.}~\bibnamefont
  {Nielsen}}\ and\ \bibinfo {author} {\bibfnamefont {I.}~\bibnamefont
  {Chuang}},\ }\href@noop {} {\emph {\bibinfo {title} {{Quantum Computation and
  Quantum Information}}}}\ (\bibinfo  {publisher} {Cambridge University
  Press},\ \bibinfo {year} {2000})\BibitemShut {NoStop}%
\bibitem [{\citenamefont {Kraus}(1983)}]{kraus}%
  \BibitemOpen
  \bibfield  {author} {\bibinfo {author} {\bibfnamefont {K.}~\bibnamefont
  {Kraus}},\ }\href@noop {} {\emph {\bibinfo {title} {{Effects and Operations:
  Fundamental Notions of Quantum Theory}}}}\ (\bibinfo  {publisher}
  {Springer},\ \bibinfo {year} {1983})\BibitemShut {NoStop}%
\bibitem [{\citenamefont {Hofheinz}\ \emph {et~al.}(2009)\citenamefont
  {Hofheinz} \emph {et~al.}}]{HofheinzNat2009}%
  \BibitemOpen
  \bibfield  {author} {\bibinfo {author} {\bibfnamefont {M.}~\bibnamefont
  {Hofheinz}} \emph {et~al.},\ }\bibfield  {title} {\bibinfo {title}
  {Synthesizing arbitrary quantum states in a superconducting resonator},\
  }\href {https://doi.org/10.1038/nature08005} {\bibfield  {journal} {\bibinfo
  {journal} {Nature}\ }\textbf {\bibinfo {volume} {459}},\ \bibinfo {pages}
  {546} (\bibinfo {year} {2009})}\BibitemShut {NoStop}%
\bibitem [{\citenamefont {Leghtas}\ \emph {et~al.}(2013)\citenamefont
  {Leghtas}, \citenamefont {Kirchmair}, \citenamefont {Vlastakis},
  \citenamefont {Devoret}, \citenamefont {Schoelkopf},\ and\ \citenamefont
  {Mirrahimi}}]{LeghtasPhysRevA2013}%
  \BibitemOpen
  \bibfield  {author} {\bibinfo {author} {\bibfnamefont {Z.}~\bibnamefont
  {Leghtas}}, \bibinfo {author} {\bibfnamefont {G.}~\bibnamefont {Kirchmair}},
  \bibinfo {author} {\bibfnamefont {B.}~\bibnamefont {Vlastakis}}, \bibinfo
  {author} {\bibfnamefont {M.~H.}\ \bibnamefont {Devoret}}, \bibinfo {author}
  {\bibfnamefont {R.~J.}\ \bibnamefont {Schoelkopf}},\ and\ \bibinfo {author}
  {\bibfnamefont {M.}~\bibnamefont {Mirrahimi}},\ }\bibfield  {title} {\bibinfo
  {title} {Deterministic protocol for mapping a qubit to coherent state
  superpositions in a cavity},\ }\href
  {https://doi.org/10.1103/PhysRevA.87.042315} {\bibfield  {journal} {\bibinfo
  {journal} {Phys. Rev. A}\ }\textbf {\bibinfo {volume} {87}},\ \bibinfo
  {pages} {042315} (\bibinfo {year} {2013})}\BibitemShut {NoStop}%
\bibitem [{\citenamefont {Sharma}\ and\ \citenamefont
  {Strauch}(2016)}]{SharmaPhysRevA2016}%
  \BibitemOpen
  \bibfield  {author} {\bibinfo {author} {\bibfnamefont {R.}~\bibnamefont
  {Sharma}}\ and\ \bibinfo {author} {\bibfnamefont {F.~W.}\ \bibnamefont
  {Strauch}},\ }\bibfield  {title} {\bibinfo {title} {Quantum state synthesis
  of superconducting resonators},\ }\href
  {https://doi.org/10.1103/PhysRevA.93.012342} {\bibfield  {journal} {\bibinfo
  {journal} {Phys. Rev. A}\ }\textbf {\bibinfo {volume} {93}},\ \bibinfo
  {pages} {012342} (\bibinfo {year} {2016})}\BibitemShut {NoStop}%
\bibitem [{\citenamefont {Wu}\ \emph {et~al.}(2017)\citenamefont {Wu},
  \citenamefont {Hou}, \citenamefont {Zhong}, \citenamefont {Yuan},
  \citenamefont {Xiang}, \citenamefont {Li},\ and\ \citenamefont {Guo}}]{Wu17}%
  \BibitemOpen
  \bibfield  {author} {\bibinfo {author} {\bibfnamefont {K.-D.}\ \bibnamefont
  {Wu}}, \bibinfo {author} {\bibfnamefont {Z.}~\bibnamefont {Hou}}, \bibinfo
  {author} {\bibfnamefont {H.-S.}\ \bibnamefont {Zhong}}, \bibinfo {author}
  {\bibfnamefont {Y.}~\bibnamefont {Yuan}}, \bibinfo {author} {\bibfnamefont
  {G.-Y.}\ \bibnamefont {Xiang}}, \bibinfo {author} {\bibfnamefont {C.-F.}\
  \bibnamefont {Li}},\ and\ \bibinfo {author} {\bibfnamefont {G.-C.}\
  \bibnamefont {Guo}},\ }\bibfield  {title} {\bibinfo {title} {Experimentally
  obtaining maximal coherence via assisted distillation process},\ }\href
  {https://doi.org/10.1364/OPTICA.4.000454} {\bibfield  {journal} {\bibinfo
  {journal} {Optica}\ }\textbf {\bibinfo {volume} {4}},\ \bibinfo {pages} {454}
  (\bibinfo {year} {2017})}\BibitemShut {NoStop}%
\bibitem [{\citenamefont {Monz}\ \emph {et~al.}(2011)\citenamefont {Monz},
  \citenamefont {Schindler}, \citenamefont {Barreiro}, \citenamefont {Chwalla},
  \citenamefont {Nigg}, \citenamefont {Coish}, \citenamefont {Harlander},
  \citenamefont {H\"ansel}, \citenamefont {Hennrich},\ and\ \citenamefont
  {Blatt}}]{MonzPRL2011}%
  \BibitemOpen
  \bibfield  {author} {\bibinfo {author} {\bibfnamefont {T.}~\bibnamefont
  {Monz}}, \bibinfo {author} {\bibfnamefont {P.}~\bibnamefont {Schindler}},
  \bibinfo {author} {\bibfnamefont {J.~T.}\ \bibnamefont {Barreiro}}, \bibinfo
  {author} {\bibfnamefont {M.}~\bibnamefont {Chwalla}}, \bibinfo {author}
  {\bibfnamefont {D.}~\bibnamefont {Nigg}}, \bibinfo {author} {\bibfnamefont
  {W.~A.}\ \bibnamefont {Coish}}, \bibinfo {author} {\bibfnamefont
  {M.}~\bibnamefont {Harlander}}, \bibinfo {author} {\bibfnamefont
  {W.}~\bibnamefont {H\"ansel}}, \bibinfo {author} {\bibfnamefont
  {M.}~\bibnamefont {Hennrich}},\ and\ \bibinfo {author} {\bibfnamefont
  {R.}~\bibnamefont {Blatt}},\ }\bibfield  {title} {\bibinfo {title} {14-qubit
  entanglement: Creation and coherence},\ }\href
  {https://doi.org/10.1103/PhysRevLett.106.130506} {\bibfield  {journal}
  {\bibinfo  {journal} {Phys. Rev. Lett.}\ }\textbf {\bibinfo {volume} {106}},\
  \bibinfo {pages} {130506} (\bibinfo {year} {2011})}\BibitemShut {NoStop}%
\bibitem [{\citenamefont {Friis}\ \emph {et~al.}(2018)\citenamefont {Friis},
  \citenamefont {Marty}, \citenamefont {Maier}, \citenamefont {Hempel},
  \citenamefont {Holz\"apfel}, \citenamefont {Jurcevic}, \citenamefont
  {Plenio}, \citenamefont {Huber}, \citenamefont {Roos}, \citenamefont
  {Blatt},\ and\ \citenamefont {Lanyon}}]{FriisPRX2018}%
  \BibitemOpen
  \bibfield  {author} {\bibinfo {author} {\bibfnamefont {N.}~\bibnamefont
  {Friis}}, \bibinfo {author} {\bibfnamefont {O.}~\bibnamefont {Marty}},
  \bibinfo {author} {\bibfnamefont {C.}~\bibnamefont {Maier}}, \bibinfo
  {author} {\bibfnamefont {C.}~\bibnamefont {Hempel}}, \bibinfo {author}
  {\bibfnamefont {M.}~\bibnamefont {Holz\"apfel}}, \bibinfo {author}
  {\bibfnamefont {P.}~\bibnamefont {Jurcevic}}, \bibinfo {author}
  {\bibfnamefont {M.~B.}\ \bibnamefont {Plenio}}, \bibinfo {author}
  {\bibfnamefont {M.}~\bibnamefont {Huber}}, \bibinfo {author} {\bibfnamefont
  {C.}~\bibnamefont {Roos}}, \bibinfo {author} {\bibfnamefont {R.}~\bibnamefont
  {Blatt}},\ and\ \bibinfo {author} {\bibfnamefont {B.}~\bibnamefont
  {Lanyon}},\ }\bibfield  {title} {\bibinfo {title} {Observation of entangled
  states of a fully controlled 20-qubit system},\ }\href
  {https://doi.org/10.1103/PhysRevX.8.021012} {\bibfield  {journal} {\bibinfo
  {journal} {Phys. Rev. X}\ }\textbf {\bibinfo {volume} {8}},\ \bibinfo {pages}
  {021012} (\bibinfo {year} {2018})}\BibitemShut {NoStop}%
\bibitem [{\citenamefont {Arute}\ \emph {et~al.}(2019)\citenamefont {Arute}
  \emph {et~al.}}]{Arute2019}%
  \BibitemOpen
  \bibfield  {author} {\bibinfo {author} {\bibfnamefont {F.}~\bibnamefont
  {Arute}} \emph {et~al.},\ }\bibfield  {title} {\bibinfo {title} {Quantum
  supremacy using a programmable superconducting processor},\ }\href
  {https://doi.org/10.1038/s41586-019-1666-5} {\bibfield  {journal} {\bibinfo
  {journal} {Nature}\ }\textbf {\bibinfo {volume} {574}} (\bibinfo {year}
  {2019})}\BibitemShut {NoStop}%
\bibitem [{\citenamefont {Wang}\ \emph {et~al.}(2019)\citenamefont {Wang},
  \citenamefont {Qin}, \citenamefont {Ding}, \citenamefont {Chen},
  \citenamefont {Chen}, \citenamefont {You}, \citenamefont {He}, \citenamefont
  {Jiang}, \citenamefont {You}, \citenamefont {Wang}, \citenamefont
  {Schneider}, \citenamefont {Renema}, \citenamefont {H\"ofling}, \citenamefont
  {Lu},\ and\ \citenamefont {Pan}}]{WangPRL2019}%
  \BibitemOpen
  \bibfield  {author} {\bibinfo {author} {\bibfnamefont {H.}~\bibnamefont
  {Wang}}, \bibinfo {author} {\bibfnamefont {J.}~\bibnamefont {Qin}}, \bibinfo
  {author} {\bibfnamefont {X.}~\bibnamefont {Ding}}, \bibinfo {author}
  {\bibfnamefont {M.-C.}\ \bibnamefont {Chen}}, \bibinfo {author}
  {\bibfnamefont {S.}~\bibnamefont {Chen}}, \bibinfo {author} {\bibfnamefont
  {X.}~\bibnamefont {You}}, \bibinfo {author} {\bibfnamefont {Y.-M.}\
  \bibnamefont {He}}, \bibinfo {author} {\bibfnamefont {X.}~\bibnamefont
  {Jiang}}, \bibinfo {author} {\bibfnamefont {L.}~\bibnamefont {You}}, \bibinfo
  {author} {\bibfnamefont {Z.}~\bibnamefont {Wang}}, \bibinfo {author}
  {\bibfnamefont {C.}~\bibnamefont {Schneider}}, \bibinfo {author}
  {\bibfnamefont {J.~J.}\ \bibnamefont {Renema}}, \bibinfo {author}
  {\bibfnamefont {S.}~\bibnamefont {H\"ofling}}, \bibinfo {author}
  {\bibfnamefont {C.-Y.}\ \bibnamefont {Lu}},\ and\ \bibinfo {author}
  {\bibfnamefont {J.-W.}\ \bibnamefont {Pan}},\ }\bibfield  {title} {\bibinfo
  {title} {Boson sampling with 20 input photons and a 60-mode interferometer in
  a $1{0}^{14}$-dimensional hilbert space},\ }\href
  {https://doi.org/10.1103/PhysRevLett.123.250503} {\bibfield  {journal}
  {\bibinfo  {journal} {Phys. Rev. Lett.}\ }\textbf {\bibinfo {volume} {123}},\
  \bibinfo {pages} {250503} (\bibinfo {year} {2019})}\BibitemShut {NoStop}%
\end{thebibliography}%

\end{document}